\documentclass[twocolumn]{autart}    % Enable this line and disable the
\usepackage{graphicx}          % Include this line if your
\usepackage{color}
\usepackage{float}
\usepackage{mathrsfs}
\usepackage{amsmath}
\usepackage{amssymb}
\usepackage{graphicx}
\usepackage{wasysym}
\usepackage{esint}
\usepackage[font={small}]{caption}
\usepackage{url}% Include this line if your document contains figures,
\usepackage{subfigure}% subcaptions for subfigures
\usepackage{epsfig}% for postscript graphics files
\usepackage{subfigmat}% matrices of similar subfigures, aka small mulitples
\usepackage{tensor}
\usepackage{amsfonts}
\usepackage{mathrsfs}
\usepackage{graphicx}
\usepackage{tikz}
\usepackage{wasysym}

\usepackage{natbib}
\newtheorem{theorem}{Theorem}
\newtheorem{lemma}[theorem]{Lemma}

\newtheorem{corollary}[theorem]{Corollary}

\newtheorem{assumption}{Assumption}
\newtheorem{remark}{Remark}
\newtheorem{definition}{Definition}

\usepackage{bigints}

\newcommand{\tikzcircle}[2][red,fill=red]{\tikz[baseline=-0.5ex]\draw[#1,radius=#2] (0,0) circle ;}%

\begin{document}

\begin{frontmatter}

%\title{Distributed Bayesian Filtering Algorithm based on the Logarithmic Opinion Pool for Dynamic Sensor Networks\thanksref{footnoteinfo}}
\title{Distributed Bayesian Filtering using Logarithmic Opinion Pool for Dynamic Sensor Networks\thanksref{footnoteinfo} }

\thanks[footnoteinfo]{S. Bandyopadhyay and S.-J. Chung were supported in part by the AFOSR grant (FA95501210193) and the NSF grant (IIS-1253758).
%This paper was not presented at any IFAC meeting.
}

\author[JPL]{Saptarshi Bandyopadhyay}\ead{Saptarshi.Bandyopadhyay@jpl.nasa.gov},    % Add the
\author[Caltech]{Soon-Jo Chung\thanksref{correspondingauthor}}\ead{sjchung@caltech.edu},               % e-mail address
%\author[Baiae]{Publius Maro Vergilius}\ead{vergilius@culture.ir}  % (ead) as shown

\address[JPL]{Jet Propulsion Laboratory, California Institute of Technology, Pasadena, CA 91109, USA}
\address[Caltech]{Graduate Aerospace Laboratories, California Institute of Technology, Pasadena, CA 91125, USA}
%\address[UIUC]{Department of Aerospace Engineering and Coordinated Science Laboratory, \\ University of Illinois at Urbana-Champaign, Urbana, IL 61801, USA}  % Please supply
%\address[Rome]{Senate House, Rome}             % full addresses
%\address[Baiae]{The White House, Baiae}        % here.
\thanks[correspondingauthor]{Corresponding author. Tel.: +1 626 395 6294.}
%\begin{keyword}                           % Five to ten keywords,
%Bayesian filtering, distributed estimation, sensor network, data fusion, logarithmic opinion pool.               % chosen from the IFAC
%\end{keyword}                             % keyword list or with the
                                          % help of the Automatica
                                          % keyword wizard

\begin{abstract}                          % Abstract of not more than 200 words.
The discrete-time Distributed Bayesian Filtering (DBF) algorithm is presented for the problem of tracking a target dynamic model using a time-varying network of heterogeneous sensing agents. In the DBF algorithm, the sensing agents combine their normalized likelihood functions in a distributed manner using the logarithmic opinion pool and the dynamic average consensus algorithm. We show that each agent's estimated likelihood function globally exponentially converges to an error ball centered on the joint likelihood function of the centralized multi-sensor Bayesian filtering algorithm. We rigorously characterize the convergence, stability, and robustness properties of the DBF algorithm. Moreover, we provide an explicit bound on the time step size of the DBF algorithm that depends on the time-scale of the target dynamics, the desired convergence error bound, and the modeling and communication error bounds.
Furthermore, the DBF algorithm for linear-Gaussian models is cast into a modified form of the Kalman information filter.
The performance and robust properties of the DBF algorithm are validated using numerical simulations.
\end{abstract}

\end{frontmatter}

\section{Introduction}
A network of time-varying, heterogeneous sensing agents could use a distributed estimation algorithm to estimate the states of the target dynamics in a distributed manner.
Potential applications include environment and pollution monitoring, analyzing communication and social networks, and tracking mobile targets on Earth or in space.
In this paper, we present a new, discrete-time distributed estimation algorithm based on the logarithmic opinion pool that guarantees bounded convergence to the Bayesian-optimal probability distribution of the states of the target dynamics.\\
\phantom{123}Discrete-time distributed estimation algorithms can be broadly classified into three categories based
on their representation of the states of the target dynamics.
Algorithms in the first category only estimate the mean and the covariance matrix of the target's states \citep{Ref:Speyer79,Ref:Borkar82,Ref:Chen02,Ref:Tomlin08,Ref:Saber09,Ref:Battistelli15,Ref:Rashedi16communication}.
These algorithms usually deal with linearized target dynamics and
measurement models, and also neglect information
captured by the higher-order moments of the estimated probability distribution
of the target's states.
The second category aims to reach an agreement across
the sensor network over a discrete set of hypotheses about the states
of the target \citep{Ref:Pavlin10,Ref:Jadbabaie12,Ref:Nedic14_Cesar}.
Although these algorithms use the entire information in the estimated
probability distribution of the target's states, they are only applicable
in cases where the target's states can be represented by a discrete
(finite) set of hypotheses. Therefore, these algorithms are not suitable
for estimation over continuous domains.\\
\phantom{123}The third category of algorithms estimates the
posterior probability distribution of the states of the target \citep{Ref:Bailey12,Ref:Ahmed13,Ref:Fraser12,Ref:Hlinka12,Ref:Hlinka14,Ref:Battistelli14,Ref:Bandyopadhyay14_ACC_BCF,Ref:Bandyopadhyay_BCF_arxiv}.
This category forms the most general class of distributed estimation
algorithms because these algorithms can be used for estimation over continuous state domains, and
can incorporate nonlinear target dynamics, heterogeneous nonlinear measurement models, and non-Gaussian uncertainties.
These algorithms also use the entire information (i.e., not
just the mean and the covariance matrix) in the estimated probability
distribution of the target's states. In light of these advantages,
this paper focuses on the development of a distributed estimation
algorithm that belongs to this third category.\\
\phantom{123}In third-category algorithms, the agents exchange their local probability
distributions with their neighboring agents and combine them using
fusion or diffusive coupling rules to estimate the aggregate probability distribution.
Schemes for combining probability distributions in a distributed manner, like the Linear Opinion
Pool (LinOP) and the Logarithmic Opinion Pool (LogOP), were first studied in the statistics literature \citep{Ref:DeGroot74,Ref:Bacharach79,Ref:French81}.
The LogOP scheme is deemed ideal for this purpose because of its favorable properties \citep{Ref:Genest86}.
\phantom{123}We now focus on distributed estimation algorithms that use the LogOP scheme.
The first such algorithm is proposed in \citep{Ref:Bailey12}.
In particular, \citep{Ref:Ahmed13} generates information-theoretically-optimal weights for the LogOP scheme.
Combining probability distributions within the exponential family (i.e., probability distributions that can be expressed as exponential functions) is discussed in \citep{Ref:Fraser12,Ref:Hlinka12}.
In the distributed estimation algorithm presented in \citep{Ref:Battistelli14} as well as in our prior work \citep{Ref:Bandyopadhyay14_ACC_BCF,Ref:Bandyopadhyay_BCF_arxiv}, the distributed sensing agents combine their local posterior probability distributions using the consensus algorithm, where the multiple consensus loops within each time step are executed much faster than the original time steps of the Bayesian filter.\\
%recursively combine their local posterior probability distributions multiple times within each time step using the consensus algorithm, where the multiple consensus loops are executed much faster than the other steps in the estimator.
\phantom{123}Moreover, \citep{Ref:Battistelli14,Ref:Bandyopadhyay14_ACC_BCF,Ref:Bandyopadhyay_BCF_arxiv} show that each agent's estimated probability distribution of the target's states converges around the pdf that minimizes the sum of Kullback--Leibler (KL) divergences from all the posterior probability distributions of the target's states.
Similar algorithms for combining local likelihood functions using the consensus algorithm are proposed in \citep{Ref:Hlinka12,Ref:Hlinka14}.
But the number of consensus loops within each estimator time step grows very fast with the number of agents due to the convergence properties of the consensus algorithm \citep{Ref:Tsitsiklis09}.
Hence, such algorithms are not feasible if the time-scale of the target dynamics is comparatively fast.
This connection between the time-scale of the target dynamics and the time step size of the distributed estimation algorithm has not been explored in the literature.\\
\phantom{123}If all the agents are perfectly connected on a complete communication graph (i.e., each agent could communicate instantaneously with every other agent without any loss of information in the communication links), then the agents can exchange their local likelihood functions and use the centralized multi-sensor Bayesian filtering algorithm to estimate the Bayesian-optimal posterior probability distribution of the target's states.
An open question is how to design a distributed estimation algorithm for a time-varying, heterogeneous sensor network on a communication graph that is much sparser than a complete graph so that each agent's estimate converges to this Bayesian-optimal posterior probability distribution of the target's states.
%In order to accomplish this objective, Our aim is to design a distributed estimation algorithm where each agent's estimated probability distribution of the target's states converges to this posterior probability distribution.
Furthermore, we assume that the time-varying communication network topology is periodically strongly connected and each agent can only communicate once with its neighboring agents during each time instant.\\
\phantom{123}In this paper, we present the Distributed Bayesian Filtering (DBF) algorithm to address this open question.
During each time instant, the agents exchange their normalized likelihood functions with their neighboring agents only once and then combine them using our fusion rule.
Our fusion rule for combining arbitrary probability distributions relies on the LogOP scheme and the dynamic average consensus algorithm \citep{Ref:Saber04,Ref:Jadbabaie03,Ref:Tsitsiklis09,Ref:Zhu10}.
We show that after finite time instants, the estimated likelihood function of each agent converges to an error ball centered on the joint likelihood function of the centralized multi-sensor Bayesian filtering algorithm.
We also provide an explicit upper bound on the time step size of the DBF algorithm that depends on the time-scale of the target dynamics and the convergence error bound.
Moreover, we analyze the effect of communication and modeling errors on the DBF algorithm.
%We perform robustness analysis to determine the effect of modeling errors on the estimated likelihood functions.
If the target dynamics are linear-Gaussian models, we show that the DBF algorithm can be simplified to the modified (Kalman) information filter.
Finally, we show that the distributed estimation algorithms in \citep{Ref:Hlinka12,Ref:Hlinka14} are special cases of the DBF algorithm.\\
\phantom{123}Furthermore, \citep{Ref:Battistelli14} analyzed their algorithm using linear-Gaussian models while \citep{Ref:Fraser12} focused on probability distributions within the exponential family.
In contrast, we present a rigorous proof technique, which was first introduced in our prior work \citep{Ref:Bandyopadhyay14_ACC_BCF,Ref:Bandyopadhyay_BCF_arxiv}, for the LogOP scheme that is applicable for general probability distributions.\\
\phantom{123}This paper is organized as follows. Section~\ref{sec:Preliminaries-and-Problem}
presents some preliminaries and the problem statement. The LogOP scheme
and some general convergence results are presented in Section~\ref{sec:LogOP}.
The DBF algorithm and its special cases are presented in Section~\ref{sec:Distributed-Bayesian-Filtering}.
Results of numerical simulations are presented in Section~\ref{sec:Numerical-Simulations}
and the paper is concluded in Section~\ref{sec:Conclusions}.
\section{Preliminaries and Problem Statement \label{sec:Preliminaries-and-Problem}}
Let $\mathbb{N}$ and $\mathbb{R}$ represent the sets of positive integers and real numbers respectively.
The state space of the target's states $\mathcal{X}$ is a closed
set in $\mathbb{R}^{n_{x}}$, where $n_{x}$ is the dimension of the
states of the target. Let $\mathscr{X}$ be the Borel $\sigma$--algebra
for $\mathcal{X}$. A probability space is defined by the three-tuple
$\{\mathcal{X},\mathscr{X},\mathbb{P}\}$, where $\mathbb{P}$ is
a complete, $\sigma$-additive probability measure on all $\mathscr{X}$.
Let $p(\boldsymbol{x})=\frac{d\mathbb{P}(\boldsymbol{x})}{d\mu(\boldsymbol{x})}$
denote the Radon--Nikod\'{y}m density of the probability distribution
$\mathbb{P}(\boldsymbol{x})$ with respect to a measure $\mu(\boldsymbol{x})$.
If $\boldsymbol{x}\in\mathcal{X}$ is continuous and $\mu(\boldsymbol{x})$
is a Lebesgue measure, $p(\boldsymbol{x})$ is the probability density
function (pdf) \citep{Ref:Chen03}. Therefore, the probability of an
event $\mathscr{A}\in\mathscr{X}$ can be written as the Lebesgue--Stieltjes
integral $\mathbb{P}(\mathscr{A})=\int_{\mathscr{A}}p(\boldsymbol{x})\thinspace d\mu(\boldsymbol{x})$.
In this paper, we only deal with the continuous case where the function
$p(\cdot)$ represents the pdf and $\mu(\cdot)$ is the Lebesgue measure.
Let $\Phi(\mathcal{X})$ represent the set of all pdfs over the state
space $\mathcal{X}$.
The $L_{1}$ distance and the KL divergence between the pdfs $\mathcal{P},\mathcal{Q}\in\Phi(\mathcal{X})$ are denoted by $D_{L_{1}}\left(\mathcal{P},\mathcal{Q}\right) = \int_{\mathcal{X}}\left|\mathcal{P}(\boldsymbol{x})-\mathcal{Q}(\boldsymbol{x})\right|\:d\mu(\boldsymbol{x})$ and
$D_{\mathrm{KL}}\left(\mathcal{P}||\mathcal{Q}\right) = \int_{\mathcal{X}} \mathcal{P}(\boldsymbol{x}) \log\left( \tfrac{\mathcal{P}(\boldsymbol{x})}{\mathcal{Q}(\boldsymbol{x})} \right) \:d\mu(\boldsymbol{x})$
respectively. Also, $\exp{(\cdot)}$ is the natural exponential function.

\subsection{Target Dynamics and Measurement Models}
Let $\boldsymbol{x}_{k}$ represent the true states of the target
at the $k^{\textrm{th}}$ time instant, where $\boldsymbol{x}_{k}\in\mathcal{X}$
for all $k\in\mathbb{N}$. The dynamics of the target in discrete
time is given by:\vspace{-5pt}
\begin{equation}
\boldsymbol{x}_{k+1}=\boldsymbol{f}_{k}(\boldsymbol{x}_{k},\boldsymbol{w}_{k},\Delta)\thinspace,\thinspace\forall k\in\mathbb{N}\thinspace,\label{eq:sys_mod}
\end{equation}
where $\boldsymbol{f}_{k}:\mathbb{R}^{n_{x}}\times \mathbb{R}^{n_{w}} \rightarrow\mathbb{R}^{n_{x}}$
is a possibly nonlinear time-varying function of the state $\boldsymbol{x}_{k}$,
$\Delta$ is the discretization time step size,
$\boldsymbol{w}_{k}$ is an independent and identically distributed (i.i.d.) process noise, and $n_{w}$ is the dimension of the process
noise vector, respectively.\\
\phantom{123}Consider a network of $N$ heterogeneous sensing agents simultaneously
tracking (\ref{eq:sys_mod}). Let $\boldsymbol{y}_{k}^{i}$ denote the
measurement taken by the $i^{\textrm{th}}$ agent at the $k^{\textrm{th}}$
time instant:\vspace{-5pt}
\begin{equation}
\boldsymbol{y}_{k}^{i}=\boldsymbol{h}_{k}^{i}(\boldsymbol{x}_{k}, \boldsymbol{v}_{k}^{i}),\thinspace\forall i\in\mathcal{V}=\{1,\ldots,N\}\thinspace,\thinspace\forall k\in\mathbb{N}\thinspace,\label{eq:mes_mod_con}
\end{equation}
where $\boldsymbol{h}_{k}^{i}:\mathbb{R}^{n_{x}}\times \mathbb{R}^{n_{vi}} \rightarrow\mathbb{R}^{n_{yi}}$
is a possibly nonlinear time-varying function of the state $\boldsymbol{x}_{k}$
and an i.i.d. measurement noise $\boldsymbol{v}_{k}^{i}$, where $n_{yi}$
and $n_{vi}$ are dimensions of the measurement and measurement noise
vectors respectively. The measurements are conditionally
independent given the target's states. We assume that the target dynamics (\ref{eq:sys_mod}) and measurement models (\ref{eq:mes_mod_con}) are known.
\subsection{Bayesian Filtering Algorithm \label{sub:Standard-Bayesian-Filtering}}
Each agent uses the Bayesian filtering algorithm to estimate
the pdf of the states of the target \citep{Ref:Pearl88,Ref:Chen03}.
Let $\boldsymbol{x}_{k|k-1}$ and $\boldsymbol{x}_{k|k}$ represent
the predicted and updated states of the target at the $k^{\textrm{th}}$
time instant. Let the pdfs $\mathcal{S}_{k}^{i}=p(\boldsymbol{x}_{k|k-1})\in\Phi(\mathcal{X})$
and $\mathcal{W}_{k}^{i}=p(\boldsymbol{x}_{k|k}) = p(\boldsymbol{x}_{k|k-1} | \boldsymbol{y}_{k}^{i}) \in\Phi(\mathcal{X})$
denote the $i^{\textrm{th}}$ agent's prior and posterior pdfs of
the target's states at the $k^{\textrm{th}}$ time instant.\\
\phantom{123}During the prediction step, the prior pdf $\mathcal{S}_{k}^{i} = p(\boldsymbol{x}_{k|k-1})$
is obtained from the previous posterior pdf $\mathcal{W}_{k-1}^{i} = p(\boldsymbol{x}_{k-1|k-1})$
using the Chapman--Kolmogorov
equation \citep{Ref:Chen03}:\vspace{-5pt}
\begin{align}
\mathcal{S}_{k}^{i} = \int_{\mathcal{X}} p(\boldsymbol{x}_{k|k-1}|\boldsymbol{x}_{k-1|k-1}) \thinspace \mathcal{W}_{k-1}^{i} \thinspace d\mu(\boldsymbol{x}_{k-1|k-1}),\label{eq:predict_stage}
\end{align}
where the probabilistic model of the state evolution   \\
$p(\boldsymbol{x}_{k|k-1}|\boldsymbol{x}_{k-1|k-1})$
is obtained from the known target dynamics model (\ref{eq:sys_mod}).
We assume that the prior pdf is available at the start of the estimation
process.\\
\phantom{123}The new measurement $\boldsymbol{y}_{k}^{i}$ is used to compute the
posterior pdf $\mathcal{W}_{k}^{i} =p(\boldsymbol{x}_{k|k}) = p(\boldsymbol{x}_{k|k-1} | \boldsymbol{y}_{k}^{i})$ during
the update step using the Bayes' rule \citep{Ref:Chen03}:\vspace{-5pt}
\begin{align}
\mathcal{W}_{k}^{i} & =\frac{p(\boldsymbol{y}_{k}^{i}|\boldsymbol{x}_{k|k-1})\thinspace \mathcal{S}_{k}^{i}}{\int_{\mathcal{X}}p(\boldsymbol{y}_{k}^{i}|\boldsymbol{x}_{k|k-1})\thinspace \mathcal{S}_{k}^{i} \thinspace d\mu(\boldsymbol{x}_{k|k-1})}\thinspace.\label{eq:update_stage}
\end{align}
The likelihood function $p(\boldsymbol{y}_{k}^{i}|\boldsymbol{x}_{k|k-1})$
is obtained from the $i^{\textrm{th}}$ agent's known measurement model (\ref{eq:mes_mod_con}).
Let the pdf $\mathcal{L}_{k}^{i} =\frac{p(\boldsymbol{y}_{k}^{i}|\boldsymbol{x}_{k|k-1})}{\int_{\mathcal{X}}p(\boldsymbol{y}_{k}^{i}|\boldsymbol{x}_{k|k-1})\thinspace d\mu(\boldsymbol{x}_{k|k-1})} \in\Phi(\mathcal{X})$ represent the normalized likelihood function.
%, i.e.,\vspace{-5pt}
%\begin{equation}
%\mathcal{L}_{k}^{i}=\frac{p(\boldsymbol{y}_{k}^{i}|\boldsymbol{x}_{k|k-1})}{\int_{\mathcal{X}}p(\boldsymbol{y}_{k}^{i}|\boldsymbol{x}_{k|k-1})\thinspace d\mu(\boldsymbol{x}_{k|k-1})}\thinspace.\label{eq:normalized_likelihood_func}
%\end{equation}
Therefore, (\ref{eq:update_stage}) is equivalent to $\mathcal{W}_{k}^{i} = \frac{\mathcal{L}_{k}^{i} \mathcal{S}_{k}^{i}}{\int_{\mathcal{X}} \mathcal{L}_{k}^{i} \mathcal{S}_{k}^{i} \thinspace d\mu(\boldsymbol{x}_{k|k-1})}$.

If all the sensing agents are connected on a complete communication
graph, the agents can exchange their likelihood functions.
Each agent can use the centralized multi-sensor Bayesian filtering algorithm to
compute the centralized posterior pdf of the target's states $\mathcal{W}_{k}^{C,i}=p(\boldsymbol{x}_{k|k}) = p(\boldsymbol{x}_{k|k-1} | \boldsymbol{y}_{k}^{1},\ldots, \boldsymbol{y}_{k}^{N}) \in\Phi(\mathcal{X})$
%. For this centralized
%Bayesian filter, the posterior pdf $\mathcal{W}_{k}^{C,i}=p(\boldsymbol{x}_{k|k})\in\Phi(\mathcal{X})$ is obtained
using the Bayes' rule \citep{Ref:Khatib08}:\vspace{-5pt}
\begin{align}
\mathcal{W}_{k}^{C,i} & =\frac{\mathcal{L}_{k}^{C} \thinspace \mathcal{S}_{k}^{i}}{\int_{\mathcal{X}}  \mathcal{L}_{k}^{C} \thinspace \mathcal{S}_{k}^{i}\thinspace d\mu(\boldsymbol{x}_{k|k-1})}\thinspace, \label{eq:update_stage_multisensor}
%\textrm{where} \quad \mathcal{L}_{k}^{C} &=\frac{\prod_{j=1}^{N}\mathcal{L}_{k}^{j}}{\int_{X}\prod_{j=1}^{N}\mathcal{L}_{k}^{j}\thinspace d\mu(\boldsymbol{x}_{k|k-1})} \thinspace . \label{eq:L-C}
\end{align}
where $\mathcal{L}_{k}^{C} = \frac{\prod_{j=1}^{N}\mathcal{L}_{k}^{j}}{\int_{X}\prod_{j=1}^{N}\mathcal{L}_{k}^{j}\thinspace d\mu(\boldsymbol{x}_{k|k-1})} $ is the normalized joint likelihood function.\\
\phantom{123}Bayesian filtering is optimal because this posterior pdf $\mathcal{W}_{k}^{C,i}$
integrates and uses all the available information expressed by probabilities \citep{Ref:Chen03}. Moreover,
an optimal state estimate with respect to any criterion can be computed
from this posterior pdf $\mathcal{W}_{k}^{C,i}$. The minimum mean-square
error (MMSE) estimate and the maximum a posteriori (MAP) estimate
are given by
$\hat{\boldsymbol{x}}_{k|k}^{MMSE} =\int_{\mathcal{X}}\boldsymbol{x}\thinspace\mathcal{W}_{k}^{C,i}\thinspace d\mu(\boldsymbol{x})$
and
$\hat{\boldsymbol{x}}_{k|k}^{MAP} =\textrm{arg}\max_{\boldsymbol{x}\in\mathcal{X}}\mathcal{W}_{k}^{C,i}$
respectively \citep{Ref:Arulampalam04}.
Other potential criteria for optimality, such as maximum likelihood,
minimum conditional KL divergence, and minimum free energy, are discussed
in \citep{Ref:Chen03,Ref:Arulampalam04}.
%Let the pdf $\mathcal{L}_{k}^{C}\in\Phi(\mathcal{X})$
%represent the normalized joint likelihood function of all the sensing
%agents at the $k^{\textrm{th}}$ time instant, i.e.,\vspace{-5pt}
%\begin{equation}
%\mathcal{L}_{k}^{C}=\frac{\left(\prod_{j=1}^{N}p(\boldsymbol{y}_{k}^{j}|\boldsymbol{x}_{k|k-1})\right)}{\int_{\mathcal{X}}\left(\prod_{j=1}^{N}p(\boldsymbol{y}_{k}^{j}|\boldsymbol{x}_{k|k-1})\right)\thinspace d\mu(\boldsymbol{x}_{k|k-1})}\thinspace.\label{eq:L-C}
%\end{equation}
%Therefore, the pdf $\mathcal{W}_{k}^{C,i} = \frac{\mathcal{L}_{k}^{C} \mathcal{S}_{k}^{i}}{\int_{\mathcal{X}} \mathcal{L}_{k}^{C} \mathcal{S}_{k}^{i} \thinspace d\mu(\boldsymbol{x}_{k|k-1})}$.
%The main advantage of Bayesian filters is that no approximation is
%needed during the filtering process
The main advantage of the original Bayesian filtering formulation is that no approximation
is needed during the filtering process; i.e., the complete information
about the dynamics and uncertainties of the model can be incorporated
in the filtering algorithm.
%However, Bayesian filtering is computationally expensive.
However, direct implementation of Bayesian filtering (\ref{eq:predict_stage})--(\ref{eq:update_stage}) is computationally expensive.
Practical implementation of these algorithms, in their
most general form, is achieved using particle filtering \citep{Ref:Pearl88,Ref:Arulampalam02}
and Bayesian programming \citep{Ref:Lebeltel04,Ref:Chen05}.\vspace{-3pt}
\subsection{Problem Statement \label{sub:Problem-Statement}}
\begin{figure}[t]
\begin{centering}
\includegraphics[bb=250bp 0bp 1510bp 1080bp,clip,width=2.8in]{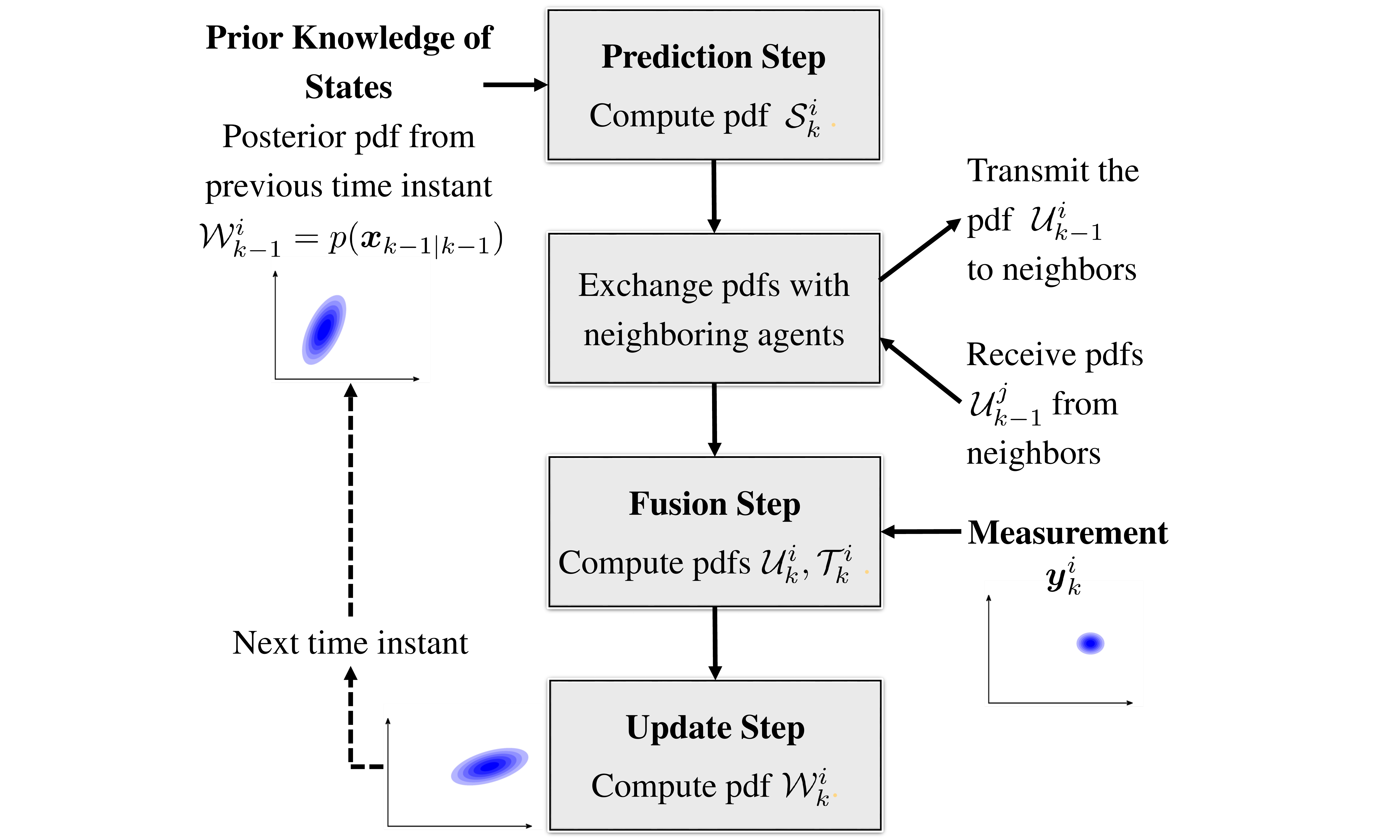}
\par\end{centering}
\caption{Flowchart of the DBF algorithm (for the $i^{\textrm{th}}$ agent at
the $k^{\textrm{th}}$ time instant) \label{fig:Flowchart-DBF}}\vspace{-5pt}
\end{figure}

Let the pdf $\mathcal{T}_{k}^{i}\in\Phi(\mathcal{X})$ denote the
estimated joint likelihood function of the $i^{\textrm{th}}$ agent at the
$k^{\textrm{th}}$ time instant. The aim is to design
a discrete-time distributed estimation algorithm, over the communication
network topology described in Section~\ref{sub:Communication-Network-Topology},
so that each agent's $\mathcal{T}_{k}^{i}$
converges to the normalized joint likelihood function $\mathcal{L}_{k}^{C}$, where the convergence error is given by:\vspace{-5pt}
\begin{align}
D_{L_{1}}\left(\mathcal{T}_{k}^{i},\mathcal{L}_{k}^{C}\right) & \leq (1+\eta)\delta\thinspace, & & \forall k\geq\kappa\thinspace,\thinspace\forall i\in\mathcal{V}\thinspace, \label{eq:problem-statement} \\
\lim_{k\rightarrow\infty} D_{L_{1}}\left(\mathcal{T}_{k}^{i},\mathcal{L}_{k}^{C}\right) & \leq \delta\thinspace, & & \forall i\in\mathcal{V}\thinspace. \label{eq:problem-statement-2}
\end{align}
where $\eta \in (0,1)$ and $\delta \in (\delta_{\mathrm{min}}, \frac{2}{1+\eta})$ denote positive constants, and $\delta_{\mathrm{min}}$ is a function of the smallest achievable time step size $\Delta_{\mathrm{min}}$, which is a practical constraint of the sensor network.\\
%where $\delta$ is the desired convergence error bound.
\phantom{123}The DBF algorithm, shown in Fig.~\ref{fig:Flowchart-DBF} and Algorithm~\ref{alg:DBF},  achieves this objective.
Note that the agents exchange their estimated pdfs with their neighboring agents only once during each time instant before the fusion step (in contrast with prior work \citep{Ref:Battistelli14,Ref:Bandyopadhyay14_ACC_BCF,Ref:Bandyopadhyay_BCF_arxiv}).
%After the prediction step, the agent's exchange their
%estimated pdfs with their neighboring agents to estimate the joint
%likelihood function. Then the agents compute the posterior pdf during
%the update step. The pseudo-code of the DBF algorithm is presented
%in Algorithm~\ref{alg:DBF}.
\subsection{Communication Network Topology \label{sub:Communication-Network-Topology}}
The time-varying communication network topology of the sensor network
is denoted by the directed graph $\mathcal{G}_{k}=(\mathcal{V},\mathcal{E}_{k})$.
%with the edge set $\mathcal{E}_{k}\subset\mathcal{V}\times\mathcal{V}$
The edge $(i,j)\in\mathcal{E}_{k}$ if and only if the $i^{\textrm{th}}$
agent receives information from the $j^{\textrm{th}}$ agent at the
$k^{\textrm{th}}$ time instant. The inclusive neighbors of the $i^{\textrm{th}}$ agent
%agent at $k^{\textrm{th}}$ time instant
are denoted by $\mathcal{J}_{k}^{i}=\{j\in\mathcal{V}:(i,j)\in\mathcal{E}_{k}\}\cup\{i\}$.
The matrix $\mathcal{A}_{k}\in\mathbb{R}^{N\times N}$ represents
the adjacency matrix of $\mathcal{G}_{k}$, where $\mathcal{A}_{k}[i,j]\not=0$
if and only if $j\in\mathcal{J}_{k}^{i}$.
%The following assumption
%on the communication network topology has been widely used in the
%literature \citep{Ref:Tsitsiklis09,Ref:Zhu10}.
\begin{assumption} \citep{Ref:Tsitsiklis09,Ref:Zhu10} \label{assump:comm-topology}
%\textit{(Periodic Strong Connectivity with Balanced Adjacency Matrix)}
The digraph $\mathcal{G}_{k}=(\mathcal{V},\mathcal{E}_{k})$
and its adjacency matrix $\mathcal{A}_{k}$ satisfy the following
properties: \\ (i) There exists some positive integer $\mathfrak{b} \in \mathbb{N}$
such that the directed graph
$\left(\mathcal{V},\mathcal{E}_{k}\cup\mathcal{E}_{k+1}\cup\ldots\cup\mathcal{E}_{k+\mathfrak{b}-1}\right)$
is strongly connected for all time instants $k\in\mathbb{N}$. \\
%If $\mathcal{G}_{k}$ is strongly connected for all time
%instants, then $\mathfrak{b}=1$. \\
(ii) The matrix $\mathcal{A}_{k}$ is doubly stochastic, i.e., $\boldsymbol{1}^{T}\mathcal{A}_{k}=\boldsymbol{1}^{T}$
and $\mathcal{A}_{k}\boldsymbol{1}=\boldsymbol{1}$ for all $k\in\mathbb{N}$, where $\mathbf{1}=[1,1,\ldots,1]^{T}$. \\
(iii) The matrix product $\mathcal{A}_{k,k+\mathfrak{b}-1}$ is defined as $\mathcal{A}_{k,k+\mathfrak{b}-1} = \left( \prod_{\tau = k}^{k+\mathfrak{b}-1} \mathcal{A}_{\tau} \right)$.
%\begin{equation}
%\mathcal{A}_{k,k+\mathfrak{b}-1} = \left( \prod_{\tau = k}^{k+\mathfrak{b}-1} \mathcal{A}_{\tau} \right) \thinspace. \label{eq:A_k_b}
%\end{equation}
There exists a constant $\gamma\in(0,\frac{1}{2})$ such that each element $\mathcal{A}_{k,k+\mathfrak{b}-1}[i,j]\in[\gamma,1] \cup \{0\} $ for all $i,j\in\mathcal{V}$ and $k \in \mathbb{N}$. Therefore, the digraph $\mathcal{G}_{k}$ is periodically strongly
connected and the matrix $\mathcal{A}_{k}$ is non-degenerate and
balanced. Note that if $\mathfrak{b} = 1$, the digraph $\mathcal{G}_{k}$ is strongly
connected at all time instants $k\in\mathbb{N}$.\vspace{-6pt}\end{assumption}

\section{Logarithmic Opinion Pool and Convergence\label{sec:LogOP}}
\begin{figure}[!h]
\begin{centering}
\begin{tabular}{cc}
\includegraphics[width=1.2in]{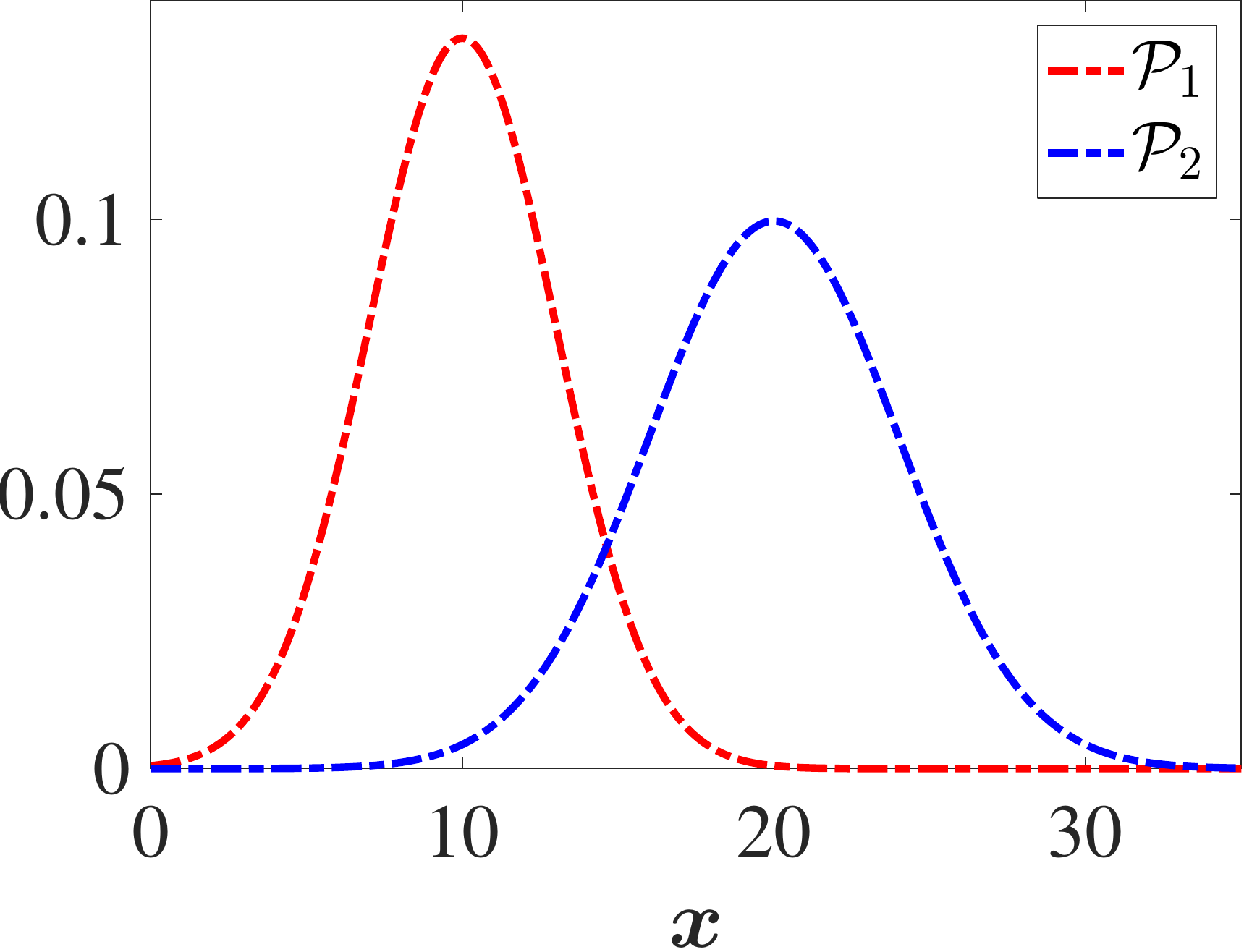} & \includegraphics[width=1.2in]{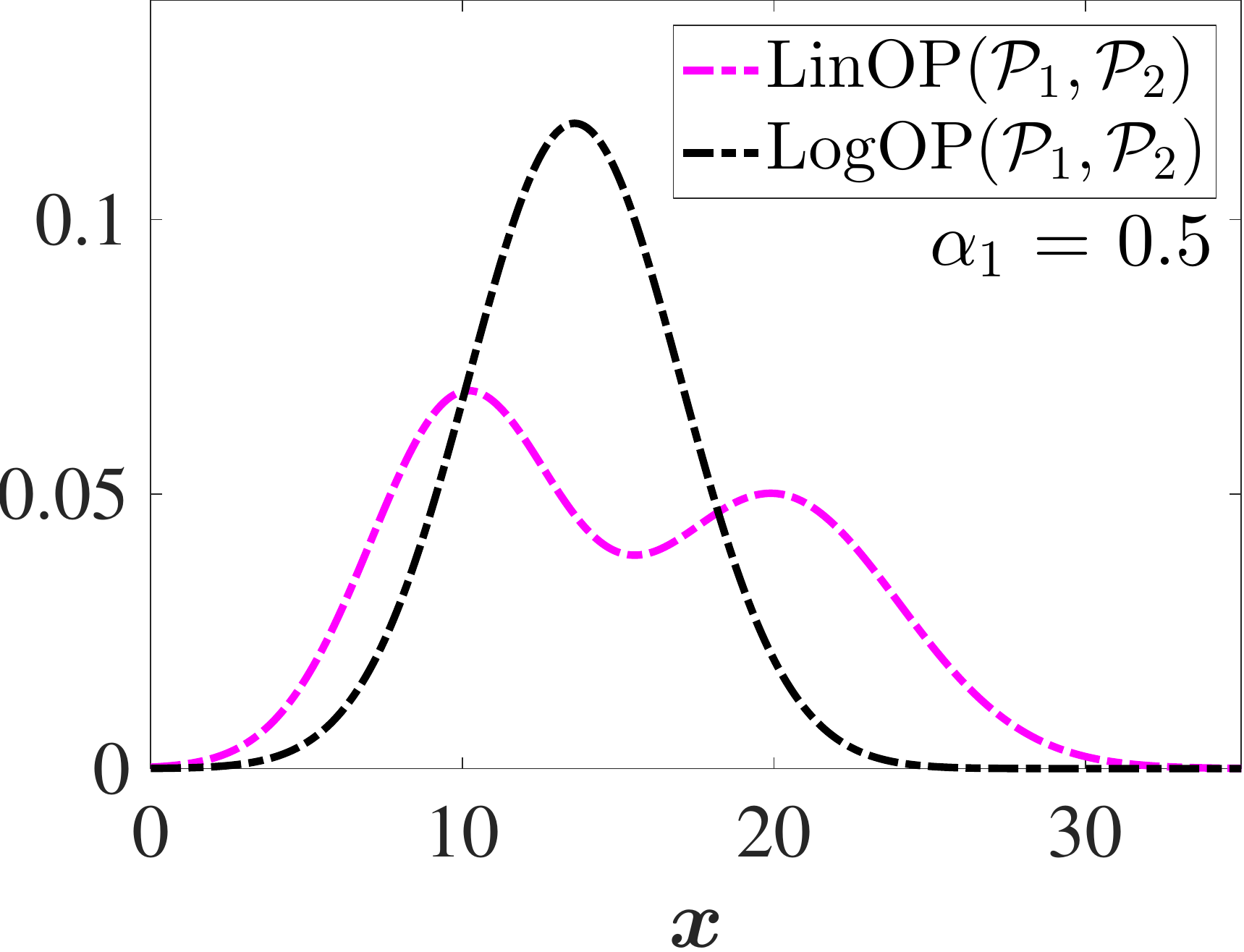}\tabularnewline
(a) & (b)\tabularnewline
\includegraphics[width=1.2in]{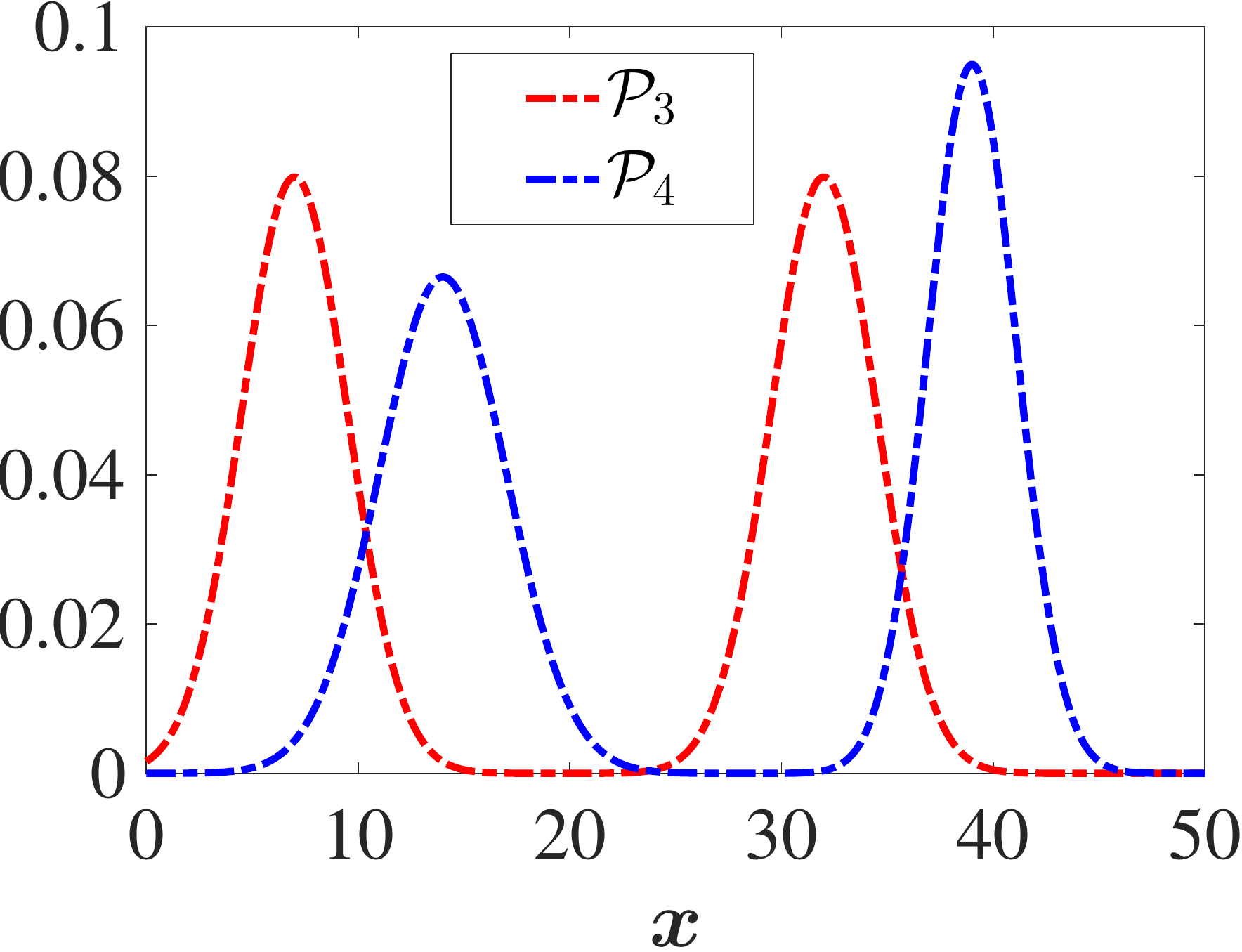} & \includegraphics[width=1.2in]{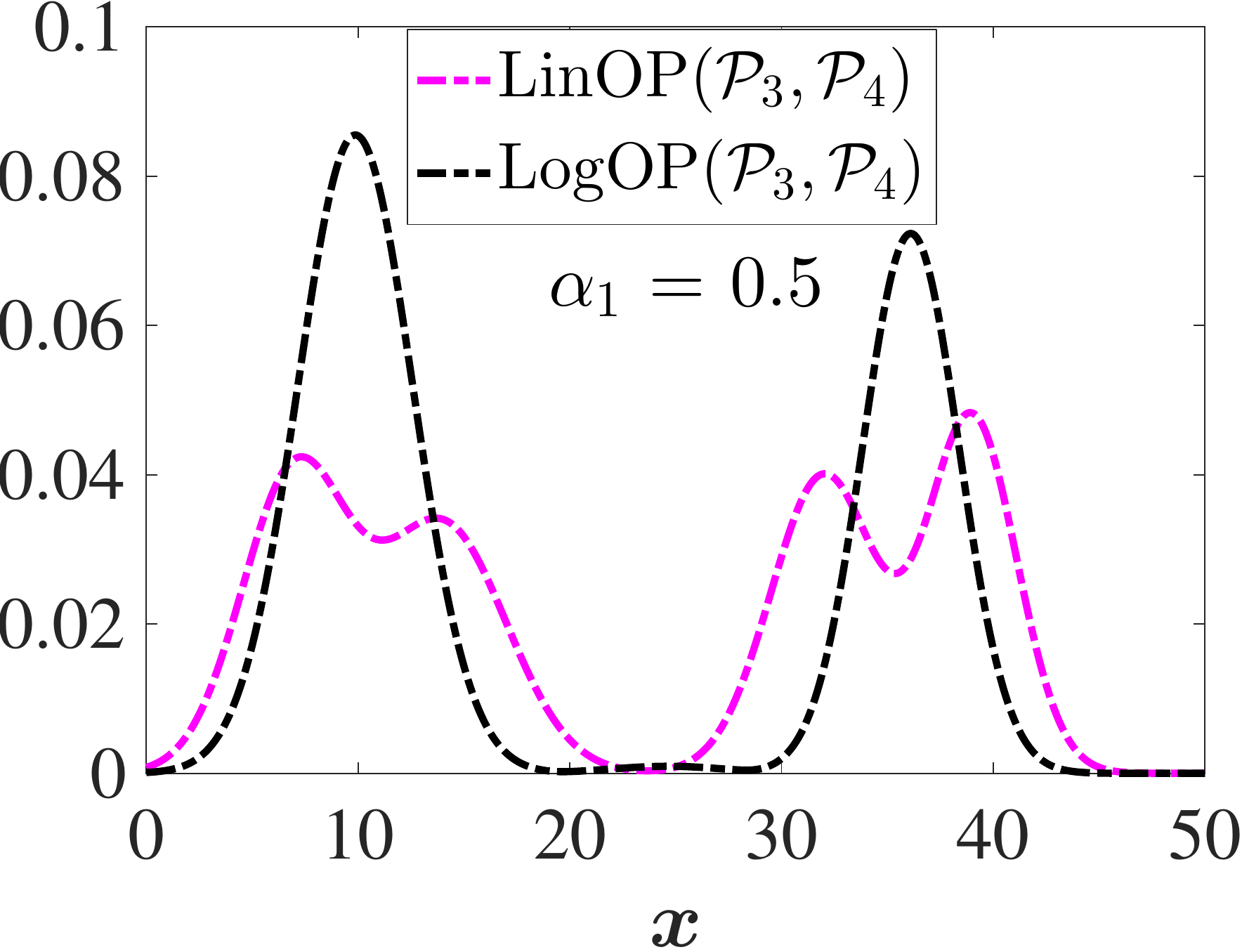}\tabularnewline
(c) & (d)\tabularnewline
\end{tabular}
\par\end{centering}
\caption{ The pdfs in (a) and (c) are combined using LinOP and LogOP in (b) and (d). Note that the LogOP solution preserves the modal nature of the original pdfs. \label{fig:Demo-combine-pdfs} }
\end{figure}

%In this section, we first state the LogOP scheme for combining probability
%distributions and then present some convergence results.
Let the pdf $\mathcal{P}_{k}^{i}\in\Phi(\mathcal{X})$ denote the
$i^{\textrm{th}}$ agent's pdf at the $k^{\textrm{th}}$ time instant.
The LinOP and LogOP schemes for combining the pdfs $\mathcal{P}_{k}^{i}$ are given by \citep{Ref:Bacharach79}:\vspace{-5pt}
\begin{align}
\mathcal{P}_{k}^{\mathrm{LinOP}}(\boldsymbol{x}) & =\sum_{i=1}^{N}\alpha_{k}^{i}\mathcal{P}_{k}^{i}(\boldsymbol{x})\thinspace,\label{eq:LinOP} \\
\mathcal{P}_{k}^{\mathrm{LogOP}}(\boldsymbol{x}) & =\frac{\Pi_{i=1}^{N}\left(\mathcal{P}_{k}^{i}(\boldsymbol{x})\right)^{\alpha_{k}^{i}}}{\int_{\mathcal{X}}\Pi_{i=1}^{N}\left(\mathcal{P}_{k}^{i}(\bar{\boldsymbol{x}})\right)^{\alpha_{k}^{i}}\thinspace d\mu(\bar{\boldsymbol{x}})}\thinspace,\label{eq:LogOP}
\end{align}
where the weights $\alpha_{k}^{i}$ are such that $\sum_{i=1}^{N}\alpha_{k}^{i}=1$
and the integral in the denominator of (\ref{eq:LogOP}) is finite.
Thus, the combined pdf obtained using LinOP and LogOP gives the weighted
algebraic and geometric averages of the individual pdfs respectively.
As shown in Fig.~\ref{fig:Demo-combine-pdfs}, the combined pdf obtained
using LogOP typically preserves the multimodal or unimodal nature
of the original individual pdfs \citep{Ref:Genest86}. The most compelling
reason for using the LogOP scheme is that it is externally Bayesian;
i.e., the LogOP combination step commutes with the process of updating
the pdfs by multiplying with a commonly agreed likelihood pdf $\mathcal{L}_{k}\in\Phi(\mathcal{X})$:\vspace{-5pt}
\begin{align*}
& \frac{\mathcal{L}_{k}\thinspace\mathcal{P}_{k}^{\mathrm{LogOP}}}{\int_{\mathcal{X}}\mathcal{L}_{k}\thinspace\mathcal{P}_{k}^{\mathrm{LogOP}} d\mu(\tilde{\boldsymbol{x}})} \! = \! \frac{\Pi_{i=1}^{N}\left(\frac{\mathcal{L}_{k}\thinspace\mathcal{P}_{k}^{i}}{\int_{\mathcal{X}}\mathcal{L}_{k}\thinspace\mathcal{P}_{k}^{i}\thinspace d\mu(\bar{\boldsymbol{x}})}\right)^{\alpha_{k}^{i}}}{\int_{\mathcal{X}}\Pi_{i=1}^{N}\left(\frac{\mathcal{L}_{k}\thinspace\mathcal{P}_{k}^{i}}{\int_{\mathcal{X}}\mathcal{L}_{k}\thinspace\mathcal{P}_{k}^{i}\thinspace d\mu(\bar{\boldsymbol{x}})}\right)^{\alpha_{k}^{i}} \! d\mu(\tilde{\boldsymbol{x}})}.
\end{align*}
Therefore, the LogOP scheme is ideal for combining pdfs in distributed
estimation algorithms.
Due to the multiplicative nature of the LogOP scheme, each agent has
veto power \citep{Ref:Genest86}. That is, if $\mathcal{P}_{k}^{i}(\boldsymbol{x})=0$
for some $\boldsymbol{x}\in\mathcal{X}$ and some agent $i\in\mathcal{V}$
with $\alpha_{k}^{i}>0$, then $\mathcal{P}_{k}^{\mathrm{LogOP}}(\boldsymbol{x})=0$
in the combined pdf irrespective of the pdfs of the other agents.
In order to avoid this veto condition, we enforce the following assumption
which has been used in the literature.
\begin{assumption} \citep{Ref:Nedic14_Cesar,Ref:Genest86} \label{assump:nonnegative_pdf} \textit{(Nonzero
Probability Property)} In this paper, all pdfs are strictly positive
everywhere in the closed set $\mathcal{X}$.
%Moreover, all pdfs are
%upper bounded by some large value. Note that the actual lower and
%upper bounds are not used in the convergence analysis.
 \end{assumption}
%We actually need the pdfs to be positive only on the support of the
%probability distribution of the target's states. Since we do not know
%this support beforehand, therefore we enforce the pdfs to be positive
%everywhere in $\mathcal{X}$.
In order to analyze the LogOP scheme with general probability distributions that satisfy Assumption~\ref{assump:nonnegative_pdf},
we use the following functions.\vspace{-3pt}
\begin{definition} \label{def:H-function}
%\textit{(Functions for Linearizing LogOP)}
Under Assumption \ref{assump:nonnegative_pdf}, for any constant $\boldsymbol{\psi}\in\mathcal{X}$, we have $\mathcal{P}_{k}^{i}(\boldsymbol{\psi})>0, \thinspace \forall i\in\mathcal{V}$ and $\mathcal{P}_{k}^{\mathrm{LogOP}}(\boldsymbol{\psi})>0$.
Using simple algebraic manipulation of (\ref{eq:LogOP}), we get \citep{Ref:Gilardoni93}:\vspace{-5pt}
\begin{align}
\mathscr{P}_{k}^{\mathrm{LogOP}}(\boldsymbol{x}) & \!:= \! \log\left[\frac{\mathcal{P}_{k}^{\mathrm{LogOP}}(\boldsymbol{x})}{\mathcal{P}_{k}^{\mathrm{LogOP}}(\boldsymbol{\psi})}\right] \!= \! \sum_{i=1}^{N} \! \alpha_{k}^{i}\mathscr{P}_{k}^{i}(\boldsymbol{x})\thinspace,\\
\textrm{where }\mathscr{P}_{k}^{i}(\boldsymbol{x}) & :=\log\left[\frac{\mathcal{P}_{k}^{i}(\boldsymbol{x})}{\mathcal{P}_{k}^{i}(\boldsymbol{\psi})}\right]\thinspace,\thinspace\forall i\in\mathcal{V}\thinspace.\label{eq:mathscr_S}
%\frac{\mathcal{P}_{k}^{\mathrm{LogOP}}(\boldsymbol{x})}{\mathcal{P}_{k}^{\mathrm{LogOP}}(\boldsymbol{\psi})} & =\frac{\left(\frac{\Pi_{i=1}^{N}\left(\mathcal{P}_{k}^{i}(\boldsymbol{x})\right)^{\alpha_{k}^{i}}}{\int_{\mathcal{X}}\Pi_{i=1}^{N}\left(\mathcal{P}_{k}^{i}(\bar{\boldsymbol{x}})\right)^{\alpha_{k}^{i}}\thinspace d\mu(\bar{\boldsymbol{x}})}\right)}{\left(\frac{\Pi_{i=1}^{N}\left(\mathcal{P}_{k}^{i}(\boldsymbol{\psi})\right)^{\alpha_{k}^{i}}}{\int_{\mathcal{X}}\Pi_{i=1}^{N}\left(\mathcal{P}_{k}^{i}(\bar{\boldsymbol{x}})\right)^{\alpha_{k}^{i}}\thinspace d\mu(\bar{\boldsymbol{x}})}\right)}\thinspace.
\end{align}
%Under Assumption \ref{assump:nonnegative_pdf}, the functions $\mathscr{P}_{k}^{i}(\boldsymbol{x})$
%and $\mathscr{P}_{k}^{\mathrm{LogOP}}(\boldsymbol{x})$ are all well-defined
%functions.
Thus, we have represented the LogOP scheme (\ref{eq:LogOP}) as a linear equation
using these functions $\mathscr{P}_{k}^{i}(\boldsymbol{x})$ and $\mathscr{P}_{k}^{\mathrm{LogOP}}(\boldsymbol{x})$, and removed the effect of the normalizing constants.
%The actual value of the constant $\boldsymbol{\psi}$ is chosen during the convergence analysis.
 \end{definition}\vspace{-3pt}
%We now define pointwise convergence and then show that the functions
%in Definition \ref{def:H-function} can be used to prove convergence
%of their corresponding pdfs.
We now state some useful convergence results using the functions in Definition~\ref{def:H-function}. See Appendix for the proofs.
\begin{definition} \textit{(Pointwise Convergence)}
%Let the pdf $\mathcal{P}^{\star}\in\Phi(\mathcal{X})$ denote the limiting pdf.
The pdf $\mathcal{P}_{k}^{i}$ converges
pointwise to the pdf $\mathcal{P}^{\star}\in\Phi(\mathcal{X})$, if and only if $\lim_{k\rightarrow\infty}\mathcal{P}_{k}^{i}(\boldsymbol{x})=\mathcal{P}^{\star}(\boldsymbol{x})$
for all $\boldsymbol{x}\in\mathcal{X}$. \end{definition}
\begin{lemma} \label{lem:psi-exisits} If the pdfs $\mathcal{P}$, $\mathcal{Q}$ satisfy Assumption~\ref{assump:nonnegative_pdf}, then there exists $\boldsymbol{\psi}\in\mathcal{X}$ such that $\mathcal{P}(\boldsymbol{\psi})=\mathcal{Q}(\boldsymbol{\psi})$. \end{lemma}
%\emph{Proof:} See Appendix~\ref{appendix:proof-lem:psi-exisits}.  \hfill $\blacksquare$
\begin{lemma} \label{lem:conv-H-func-implies-conv-pdf}
%We define
%the function $\mathscr{P}^{\star}(\boldsymbol{x}):=\ln\left[\frac{\mathcal{P}^{\star}(\boldsymbol{x})}{\mathcal{P}^{\star}(\boldsymbol{\psi})}\right]$
%for any $\boldsymbol{\psi}\in\mathcal{X}$. Under Assumption \ref{assump:nonnegative_pdf},
If the function $\mathscr{P}_{k}^{i}$ (\ref{eq:mathscr_S})
converges pointwise to the function $\mathscr{P}^{\star}:=\log\left[\frac{\mathcal{P}^{\star}(\boldsymbol{x})}{\mathcal{P}^{\star}(\boldsymbol{\psi})}\right]$, then the
corresponding pdf $\mathcal{P}_{k}^{i}$ also converges pointwise
to the pdf $\mathcal{P}^{\star}$.
\end{lemma}\vspace{-3pt}
\begin{definition} \textit{(Convergence in TV)} The measure $\mu_{\mathcal{P}_{k}^{i}}$
is defined as the measure induced by the pdf $\mathcal{P}_{k}^{i}$
on $\mathscr{X}$, where $\mu_{\mathcal{P}_{k}^{i}}(\mathscr{A})=\int_{\mathscr{A}}\mathcal{P}_{k}^{i}\thinspace d\mu(\boldsymbol{x})$
for any event $\mathscr{A}\in\mathscr{X}$. Similarly, let $\mu_{\mathcal{P}^{\star}}$
denote the measure induced by the pdf $\mathcal{P}^{\star}$ on $\mathscr{X}$. The TV distance is defined as $\|\mu_{\mathcal{P}_{k}^{i}}-\mu_{\mathcal{P}^{\star}}\|_{\mathrm{TV}}:=\sup_{\mathscr{A}\in\mathscr{X}}|\mu_{\mathcal{P}_{k}^{i}}(\mathscr{A})-\mu_{\mathcal{P}^{\star}}(\mathscr{A})|$.
The measure $\mu_{\mathcal{P}_{k}^{i}}$ converges to the measure
$\mu_{\mathcal{P}^{\star}}$ in TV, if and only if $\|\lim_{k\rightarrow\infty}\mu_{\mathcal{P}_{k}^{i}}-\mu_{\mathcal{P}^{\star}}\|_{\mathrm{TV}}=0$. \end{definition}\vspace{-3pt}
\begin{lemma} \label{lem:convegence-TV}
%\textit{(Pointwise Convergence implies Convergence in TV)}
If the pdf $\mathcal{P}_{k}^{i}$ converges pointwise to the pdf $\mathcal{P}^{\star}$, then the measure $\mu_{\mathcal{P}_{k}^{i}}$ converges in TV to the measure $\mu_{\mathcal{P}^{\star}}$. Moreover,
$\|\mu_{\mathcal{P}_{k}^{i}}-\mu_{\mathcal{P}^{\star}}\|_{\mathrm{TV}}=\frac{1}{2}D_{L_{1}}\left(\mathcal{P}_{k}^{i},\mathcal{P}^{\star}\right)$.
\end{lemma}
%\emph{Proof:} See Appendix~\ref{appendix:proof-lem:convegence-TV}. \hfill $\blacksquare$
Another reason for using the LogOP scheme is that it minimizes the information lost during the combination process, where the information loss is measured using the KL divergence.\vspace{-3pt}
% gives the KL-divergence-minimizing
%pdf as shown in the following lemma.
\begin{lemma} \label{lem:min-KL} \citep{Ref:Battistelli14,Ref:Bandyopadhyay14_ACC_BCF}
The pdf $\mathcal{P}_{k}^{\mathrm{KL}}\in\Phi(\mathcal{X})$ that globally
minimizes the sum of KL divergences with the pdfs $\mathcal{P}_{k}^{i}$
for all agents is given by:\vspace{-5pt}
\begin{align*}
\mathcal{P}_{k}^{\mathrm{KL}} \!=\! \underset{\rho\in\Phi(\mathcal{X})}{\mathrm{arg\thinspace min}} \sum_{i=1}^{N}D_{\mathrm{KL}}\left(\rho||\mathcal{P}_{k}^{i}\right) \!=\! \frac{\prod_{i=1}^{N}\left(\mathcal{P}_{k}^{i}\right)^{\frac{1}{N}}}{\int_{\mathcal{X}}\prod_{i=1}^{N}\left(\mathcal{P}_{k}^{i}\right)^{\frac{1}{N}}d\mu(\bar{\boldsymbol{x}})}.
\end{align*}
Note that the pdf $\mathcal{P}_{k}^{\mathrm{KL}}$ is
equivalent to the pdf $\mathcal{P}_{k}^{\mathrm{LogOP}}$ (\ref{eq:LogOP})
obtained using the LogOP scheme with weights $\alpha_{k}^{i}=\frac{1}{N}$
for all agents. \vspace{-5pt}\end{lemma}

%\emph{Proof:} See our prior work \citep{Ref:Bandyopadhyay14_ACC_BCF}. \hfill $\blacksquare$

%Since the KL divergence is the measure of the information lost during the combination process,
%process of combining the pdfs $\mathcal{P}_{k}^{i}$ of all agents,
%the pdf $\mathcal{P}_{k}^{\mathrm{KL}}$ minimizes the information lost in
%this combination process.
The proof of Lemma~\ref{lem:min-KL} is given in our prior work \citep{Ref:Bandyopadhyay14_ACC_BCF}. Note that the normalized joint likelihood function $\mathcal{L}_{k}^{C}$ is also given by:
\begin{align}
& \mathcal{L}_{k}^{C} =\frac{\prod_{j=1}^{N}\mathcal{L}_{k}^{j}}{\int_{X}\prod_{j=1}^{N}\mathcal{L}_{k}^{j}\thinspace d\mu(\bar{\boldsymbol{x}})}=\frac{\left(\mathcal{L}_{k}^{\mathrm{KL}}\right)^{N}}{\int_{X}\left(\mathcal{L}_{k}^{\mathrm{KL}}\right)^{N}\thinspace d\mu(\bar{\boldsymbol{x}})}\thinspace,\label{eq:L-C}\\
& \textrm{where } \mathcal{L}_{k}^{\mathrm{KL}}=\frac{\prod_{j=1}^{N}\left(\mathcal{L}_{k}^{j}\right)^{\frac{1}{N}}}{\int_{\mathcal{X}}\prod_{j=1}^{N}\left(\mathcal{L}_{k}^{j}\right)^{\frac{1}{N}}\:d\mu(\bar{\boldsymbol{x}})}\thinspace.\label{eq:L_KL}
\end{align}
We show that the DBF algorithm also estimates the pdf $\mathcal{L}_{k}^{\mathrm{KL}}$~(\ref{eq:L_KL}) in a distributed manner.

\section{Distributed Bayesian Filtering Algorithm \label{sec:Distributed-Bayesian-Filtering}}
In this section, we present the main DBF algorithm, its convergence and robustness properties, and its extensions.
%\subsection{Main Algorithm}
We first state an assumption on the time-varying nature of the pdfs $\mathcal{L}_{k}^{i}$ for all agents that directly
link the target dynamics and measurement models with the time step size
of the distributed estimation algorithm.
\begin{assumption} \label{assump:bounded_likelihood_func}
%\textit{(Relatively Bounded Normalized Likelihood Functions)}
For any time step size $\Delta>0$, there
exists a time-invariant constant $\theta_{L}>0$ such that for all
agents $i \in \mathcal{V} = \{1, \ldots, N \}$:\vspace{-5pt}
\begin{equation}
e^{-\Delta\theta_{L}}\leq\frac{\mathcal{L}_{k}^{i}(\boldsymbol{x})}{\mathcal{L}_{k-1}^{i}(\boldsymbol{x})}\leq e^{\Delta\theta_{L}}\thinspace,\thinspace\forall\boldsymbol{x}\in\mathcal{X},\thinspace\forall k\in\mathbb{N}\thinspace.\label{eq:bounded_DeltaL}
\end{equation}
The necessary conditions for satisfying (\ref{eq:bounded_DeltaL})
are given by $D_{\mathrm{KL}}\left(\mathcal{L}_{k}^{i}||\mathcal{L}_{k-1}^{i}\right) \leq\Delta\theta_{L}$ and $D_{\mathrm{KL}}\left(\mathcal{L}_{k-1}^{i}||\mathcal{L}_{k}^{i}\right) \leq\Delta\theta_{L}$.
%A conservative $\theta_{L}$ always exists due to Assumption \ref{assump:nonnegative_pdf}.
 \end{assumption}
%Note that Assumption \ref{assump:bounded_likelihood_func} directly
%couples the target dynamics and measurement models with the time step
%of the distributed estimation algorithm. We later show that the convergence error bound also depends on the time step of the distributed estimation algorithm.
We now state the DBF algorithm, whose steps are shown in Fig.~\ref{fig:Flowchart-DBF}.
Let the pdf $\mathcal{U}_{k}^{i}\in\Phi(\mathcal{X})$ denote the estimated KL-divergence-minimizing pdf of the $i^{\textrm{th}}$ agent at the $k^{\textrm{th}}$ time instant.
The pdf $\mathcal{T}_{k}^{i}$ is defined in Section~\ref{sub:Problem-Statement}.
%and $\mathcal{T}_{k}^{i}\in\Phi(\mathcal{X})$ respectively denote
%the estimated KL-divergence-minimizing pdf and the estimated joint likelihood
%function of the $i^{\textrm{th}}$ agent at the $k^{\textrm{th}}$
%time instant.
Under Assumptions \ref{assump:comm-topology}--\ref{assump:bounded_likelihood_func}, the pseudo-code of the DBF algorithm is given in Algorithm~\ref{alg:DBF}.
\begin{algorithm}[!h]
\caption{Distributed Bayesian Filtering Algorithm \label{alg:DBF}}
\centering{}%
\resizebox{3.4in}{!}{
\begin{tabular}{ll}
\hline
1. & ($i^{\textrm{th}}$ agent's steps at $k^{\textrm{th}}$ time instant)\tabularnewline
2. & Compute prior pdf $\mathcal{S}_{k}^{i}=p(\boldsymbol{x}_{k|k-1})$
using (\ref{eq:predict_stage}).\tabularnewline
3. & Obtain local measurement $\boldsymbol{y}_{k}^{i}$.\tabularnewline
4. & Compute normalized likelihood function $\mathcal{L}_{k}^{i}$. \tabularnewline
5. & Receive pdfs $\mathcal{U}_{k-1}^{j}$ from agents $j\in\mathcal{J}_{k}^{i}$.\tabularnewline
6. &  Compute pdfs $\mathcal{U}_{k}^{i}$ and $\mathcal{T}_{k}^{i}$ as follows: \vspace{-5pt} \tabularnewline
    &  \vbox{\begin{align}
 & \mathcal{U}_{k}^{i}=\dfrac{\Lambda\thinspace\mathcal{L}_{k}^{i}\left(\mathcal{L}_{k-1}^{i}\right)^{-1}}{\int_{\mathcal{X}}\Lambda\thinspace\mathcal{L}_{k}^{i}\left(\mathcal{L}_{k-1}^{i}\right)^{-1}d\mu(\bar{\boldsymbol{x}})}, \ \ \forall k\geq2
\thinspace,\label{eq:FODAC-DBFA1}\\
 & \textrm{where }\Lambda=\prod\limits _{j\in\mathcal{J}_{k}^{i}}(\mathcal{U}_{k-1}^{j})^{\mathcal{A}_{k}[i,j]}, \textrm{ and $\mathcal{U}_{1}^{i}=\mathcal{L}_{1}^{i}$ if $k=1$.}\thinspace\nonumber  \\
 & \mathcal{T}_{k}^{i}=\frac{(\mathcal{U}_{k}^{i})^{N}}{\int_{\mathcal{X}}(\mathcal{U}_{k}^{i})^{N}d\mu(\bar{\boldsymbol{x}})}\thinspace.\label{eq:FODAC-DBFA2}
\end{align}} \vspace{-5pt} \tabularnewline
7. & Compute posterior pdf $\mathcal{W}_{k}^{i}=p(\boldsymbol{x}_{k|k})$ as follows: \vspace{-5pt} \tabularnewline
    & \vbox{\begin{align}
\mathcal{W}_{k}^{i}=p(\boldsymbol{x}_{k|k})=\begin{aligned}\frac{\mathcal{T}_{k}^{i}\thinspace\mathcal{S}_{k}^{i}}{\int_{\mathcal{X}}\mathcal{T}_{k}^{i}\thinspace\mathcal{S}_{k}^{i}\thinspace d\mu(\bar{\boldsymbol{x}})}\thinspace.\end{aligned}
\label{eq:update_stage-DBFA}
\end{align}} \vspace{-8pt} \tabularnewline
\hline
\end{tabular}
}
\end{algorithm}\vspace{-8pt}

The following theorem shows that the DBF algorithm satisfies the problem statement (\ref{eq:problem-statement})--(\ref{eq:problem-statement-2}) in Section~\ref{sub:Problem-Statement}.
Here, $\eta \in (0,1)$ and $\delta \in (\delta_{\mathrm{min}}, \frac{2}{1+\eta})$ are positive constants defined in Section~\ref{sub:Problem-Statement}, $\mathfrak{b}$ is the periodicity of the communication network topology, $\gamma \in (0,\frac{1}{2})$ is the smallest positive element in $\mathcal{A}_{k,k+\mathfrak{b}-1}$ defined in Assumption~\ref{assump:comm-topology}, and $\theta_{L}$ is defined in Assumption~\ref{assump:bounded_likelihood_func}.

\begin{theorem} \label{thm:conv-likelihood}
Under Assumptions \ref{assump:comm-topology}--\ref{assump:bounded_likelihood_func}, if all the agents execute the DBF algorithm (Algorithm~\ref{alg:DBF}) and the time step size $\Delta$ for (\ref{eq:bounded_DeltaL}) for Algorithm~\ref{alg:DBF} is defined as\vspace{-5pt}
\begin{equation}
\Delta = \frac{\left(1-\sigma_{m}\right)\log\left(\delta+1\right)}{2\mathfrak{b}N(N-1)\sqrt{N}\theta_{L}}  \thinspace, \label{eq:Delta_max}
\end{equation}
then the steady-state convergence error between the pdf $\mathcal{T}_{k}^{i}$ (\ref{eq:FODAC-DBFA2})
and the pdf $\mathcal{L}_{k}^{C}$ (\ref{eq:L-C}) is bounded by $\delta \in (\delta_{\mathrm{min}}, \frac{2}{1+\eta})$:\vspace{-5pt}
\begin{align}
&\lim_{k\rightarrow\infty} \thinspace \max_{i\in\mathcal{V}=\{1,\ldots,N\} }D_{L_{1}}\left(\mathcal{T}_{k}^{i},\mathcal{L}_{k}^{C}\right)\leq \delta \thinspace, \label{eq:T_error-ss} \\
& \delta_{\mathrm{min}} = \exp{ \left(\frac{\Delta_{\mathrm{min}} 2\mathfrak{b}N(N-1)\sqrt{N}\theta_{L}}{1-\sigma_{m}} \right)} - 1 \thinspace. \label{eq:delta_min}
\end{align}
Furthermore, the convergence error between the pdfs $\mathcal{T}_{k}^{i}$~(\ref{eq:FODAC-DBFA2}) and $\mathcal{L}_{k}^{C}$~(\ref{eq:L-C}) after $\kappa$ time instants is bounded as:\vspace{-5pt}
\begin{align}
&\max_{i\in\mathcal{V} =\{1,\ldots,N\} }D_{L_{1}}\left(\mathcal{T}_{k}^{i},\mathcal{L}_{k}^{C}\right)\leq (1+\eta)\delta \thinspace, \thinspace \forall k\geq\kappa\thinspace, \label{eq:T_error}
\end{align}
where, if $\mathfrak{D}_{1} = 2\log\left(\max_{\ell,j\in\mathcal{V}}\max_{\boldsymbol{x}\in\mathcal{X}} \frac{\mathcal{L}_{1}^{\ell}(\boldsymbol{x})}{\mathcal{L}_{1}^{j}(\boldsymbol{x})}\right)\leq \frac{\log\left(\delta+1\right)}{N^{\frac{3}{2}}}$,  $\kappa = 1$. Otherwise,\vspace{-5pt}
\begin{align}
\kappa= \left\lceil \frac{\mathfrak{b}(N-1)}{\log\sigma_{m}}\log\left(\frac{\log\left(\frac{(1+\eta)\delta+1}{\delta+1}\right)}{\log\left(\frac{e^{N^{\frac{3}{2}}\mathfrak{D}_{1}}}{\delta+1}\right)}\right)\right\rceil  +1 \thinspace.  \label{eq:kappa1}
\end{align}
Here, $\sigma_{m} = \max_{k\in \mathbb{N}} \sigma_{N-1}(\mathcal{A}_{k,k+\mathfrak{b}(N-1)-1})$, where  $\sigma_{N-1}$ denotes the second largest singular value of the matrix, and $\sigma_{m}$ is upper bounded by:\vspace{-5pt}
\begin{equation}
\sigma_{m} \leq \left(1-\frac{4(\gamma-\gamma^{N})}{(1-\gamma)}\sin^{2}\frac{\pi}{2N}\right)^{\frac{1}{2}} < 1 \thinspace. \label{eq:sigma_m_bound}
\end{equation}
The TV error between the measures induced by the pdfs $\mathcal{T}_{k}^{i}$
and $\mathcal{L}_{k}^{C}$ is bounded by:\vspace{-5pt}
\begin{align}
& \max_{i\in\mathcal{V} } \|\mu_{\mathcal{T}_{k}^{i}}-\mu_{\mathcal{L}_{k}^{C}}\|_{\mathrm{TV}}\leq \frac{(1+\eta)\delta}{2} \thinspace, \quad \forall k\geq\kappa \thinspace , \\
& \lim_{k\rightarrow\infty} \max_{i\in\mathcal{V} } \|\mu_{\mathcal{T}_{k}^{i}}-\mu_{\mathcal{L}_{k}^{C}}\|_{\mathrm{TV}} \leq \frac{\delta}{2} \thinspace .
\end{align}
\end{theorem}
\emph{Proof:} Using Definition~\ref{def:H-function},  we define $\mathscr{L}_{k}^{\mathrm{KL}}(\boldsymbol{x}) =\log\left[\frac{\mathcal{L}_{k}^{\mathrm{KL}}(\boldsymbol{x})}{\mathcal{L}_{k}^{\mathrm{KL}}(\boldsymbol{\psi})}\right]$,
$\mathscr{L}_{k}^{C}(\boldsymbol{x}) =\log\left[\frac{\mathcal{L}_{k}^{C}(\boldsymbol{x})}{\mathcal{L}_{k}^{C}(\boldsymbol{\psi})}\right]$,
$\mathscr{L}_{k}^{i}(\boldsymbol{x}) =\log\left[\frac{\mathcal{L}_{k}^{i}(\boldsymbol{x})}{\mathcal{L}_{k}^{i}(\boldsymbol{\psi})}\right]$,
$\mathscr{U}_{k}^{i}(\boldsymbol{x}) =\log\left[\frac{\mathcal{U}_{k}^{i}(\boldsymbol{x})}{\mathcal{U}_{k}^{i}(\boldsymbol{\psi})}\right]$, and $\mathscr{T}_{k}^{i}(\boldsymbol{x}) =\log\left[\frac{\mathcal{T}_{k}^{i}(\boldsymbol{x})}{\mathcal{T}_{k}^{i}(\boldsymbol{\psi})}\right]$ for all  $i\in\mathcal{V}$. Since these functions are defined for all $\boldsymbol{x} \in \mathcal{X}$, we henceforth drop the term $(\boldsymbol{x})$ for brevity.

\textbf{Step 1.} We first show that the pdf $\mathcal{U}_{k}^{i}$ (\ref{eq:FODAC-DBFA1})
converges to the pdf $\mathcal{L}_{k}^{\mathrm{KL}}$ (\ref{eq:L_KL}). Equation
(\ref{eq:FODAC-DBFA1}) can be re-written using these functions as:\vspace{-5pt}
\begin{align}
\mathscr{U}_{k}^{i} & \!=\! \begin{cases}
\mathscr{L}_{1}^{i} & \textrm{if } k=1\\
\sum_{j=1}^{N}\! \mathcal{A}_{k}[i,j]\mathscr{U}_{k-1}^{j} \!+\! \mathscr{L}_{k}^{i} \!-\! \mathscr{L}_{k-1}^{i} \!\!\! & \textrm{if } k\geq2
\end{cases},\label{eq:proof_step1}
\end{align}
because $\mathcal{A}_{k}[i,j]=0$ if $j\not\in\mathcal{J}_{k}^{i}$, as defined in Section~\ref{sub:Communication-Network-Topology}.
Since $\mathcal{A}_{k}$ is doubly stochastic, (\ref{eq:proof_step1})
satisfies the conservation property:\vspace{-5pt}
\begin{align}
& \sum_{i=1}^{N}\mathscr{U}_{k}^{i} = \sum_{i=1}^{N} \sum_{j=1}^{N}\mathcal{A}_{k}[i,j]\mathscr{U}_{k-1}^{j} +\sum_{i=1}^{N}\left(\mathscr{L}_{k}^{i}-\mathscr{L}_{k-1}^{i}\right)\thinspace,\nonumber \\
& =\sum_{i=1}^{N} \left( \sum_{j=1}^{N}\mathcal{A}_{k}[j,i] \right) \mathscr{U}_{k-1}^{i}+\sum_{i=1}^{N}\left(\mathscr{L}_{k}^{i}-\mathscr{L}_{k-1}^{i}\right)\thinspace,\nonumber \\
& =\sum_{i=1}^{N}\left(\mathscr{U}_{1}^{i}-\mathscr{L}_{1}^{i}\right)+\sum_{i=1}^{N}\mathscr{L}_{k}^{i}=N\mathscr{L}_{k}^{\mathrm{KL}}\thinspace.\label{eq:proof_step8}
\end{align}
Note that $\mathscr{L}_{k}^{\mathrm{KL}} = \frac{1}{N} \sum_{i=1}^{N}\mathscr{L}_{k}^{i}$ follows from (\ref{eq:L_KL}).
This shows that if the functions $\mathscr{U}_{k}^{i}$ converge towards each other, then they will converge to the function $\mathscr{L}_{k}^{\mathrm{KL}}$.
Let us define the error vector $\boldsymbol{e}_k$ as:\vspace{-5pt}
\begin{align}
\boldsymbol{e}_k = \left[\begin{array}{ccccc}
\mathscr{U}_{k}^{1} \!-\! \mathscr{L}_{k}^{\mathrm{KL}}, & \ldots, & \mathscr{U}_{k}^{i} \!-\! \mathscr{L}_{k}^{\mathrm{KL}}, & \ldots, & \mathscr{U}_{k}^{N} \!-\! \mathscr{L}_{k}^{\mathrm{KL}} \end{array}\right]^{T}. \nonumber
\end{align}
The evolution of the error vector $\boldsymbol{e}_k$ is given by:\vspace{-5pt}
\begin{align}
& & \boldsymbol{e}_{k} &= \mathcal{A}_{k} \boldsymbol{e}_{k-1} + \boldsymbol{\Omega}_{k,k} \thinspace, \qquad \forall k \geq 2 \thinspace,  \\
& \textrm{where} & \boldsymbol{\Omega}_{k,k} &= \left(\mathbf{I} - \frac{\boldsymbol{11}^{T}}{N} \right) \left[ \begin{smallmatrix}
\mathscr{L}_{k}^{1}-\mathscr{L}_{k-1}^{1}\\
\vdots\\
\mathscr{L}_{k}^{N}-\mathscr{L}_{k-1}^{N}
\end{smallmatrix}\right] \thinspace. \nonumber
\end{align}
The overall evolution of the error vector $\boldsymbol{e}_k$ after $\mathfrak{b} \in \mathbb{N}$ time instants is given by:\vspace{-5pt}
\begin{align}
& \boldsymbol{e}_{k+\mathfrak{b}-1}  =  \mathcal{A}_{k,k+\mathfrak{b}-1} \boldsymbol{e}_{k-1} + \boldsymbol{\Omega}_{k,k+\mathfrak{b}-1} \thinspace, \label{eq:ekb1}
\end{align}
where $\mathcal{A}_{k,k+\mathfrak{b}-1}$ is defined in Assumption~\ref{assump:comm-topology} and for $\mathfrak{b} \geq 2$:
\begin{align}
& \boldsymbol{\Omega}_{k,k+\mathfrak{b}-1} = \sum_{\tau=k}^{k+\mathfrak{b}-2}\left(\mathcal{A}_{\tau+1,k+\mathfrak{b}-1}\boldsymbol{\Omega}_{\tau,\tau}\right) +\boldsymbol{\Omega}_{k+\mathfrak{b}-1,k+\mathfrak{b}-1} \thinspace. \nonumber
\end{align}
Note that $\boldsymbol{1}^{T} \boldsymbol{e}_k = 0$ because of (\ref{eq:proof_step8}) and $\boldsymbol{1}^{T} \boldsymbol{\Omega}_{k,k+\mathfrak{b}-1} = 0$ because $\boldsymbol{1}^{T} \left(\mathbf{I} - \frac{\boldsymbol{11}^{T}}{N} \right) = 0$.
Therefore, we investigate the convergence of $\boldsymbol{e}_k$ along all directions that are orthogonal to $\boldsymbol{1}^{T}$.
{It follows from Assumption~\ref{assump:comm-topology} that the matrix  $\mathcal{A}_{k,k+\mathfrak{b}-1}$ is irreducible. Therefore, the matrix $\mathcal{A}_{k,k+\mathfrak{b}-1}$ is primitive \cite[Lemma 8.5.4, pp. 516]{Ref:Horn85} and $ | \lambda_{N-1} \left( \mathcal{A}_{k,k+\mathfrak{b}-1} \right) | < 1$, where $ \lambda_{N-1}$ denotes the second largest modulus of eigenvalues of the matrix and $|\cdot|$ represents the complex modulus.}
Let $V_{\textrm{tr}}=\left[\frac{1}{\sqrt{N}}\mathbf{1},\thinspace V_{\textrm{s}}\right]$ be the orthonormal matrix of eigenvectors of the symmetric primitive matrix $\mathcal{A}_{1,\mathfrak{b}}^T \mathcal{A}_{1,\mathfrak{b}}$. By spectral decomposition \citep{Ref:Chung12}, we get:\vspace{-5pt}
\begin{equation}
V_{\textrm{tr}}^{T} \mathcal{A}_{1,\mathfrak{b}}^T \mathcal{A}_{1,\mathfrak{b}} V_{\textrm{tr}}=\left[\begin{smallmatrix}
1 & \mathbf{0}^{1\times(N-1)}\\
\mathbf{0}^{(N-1)\times1} & V_{\textrm{s}}^{T} \mathcal{A}_{1,\mathfrak{b}}^T \mathcal{A}_{1,\mathfrak{b}} V_{\textrm{s}}
\end{smallmatrix}\right]\thinspace, \nonumber
\end{equation}
where $\frac{1}{N}\mathbf{1}^{T} \mathcal{A}_{1,\mathfrak{b}}^T \mathcal{A}_{1,\mathfrak{b}} \mathbf{1}=1$, $\frac{1}{\sqrt{N}}\mathbf{1}^{T} \mathcal{A}_{1,\mathfrak{b}}^T \mathcal{A}_{1,\mathfrak{b}} V_{\textrm{s}}=\mathbf{0}^{1\times(N-1)}$, and $V_{\textrm{s}}^{T} \mathcal{A}_{1,\mathfrak{b}}^T \mathcal{A}_{1,\mathfrak{b}} \mathbf{1}\frac{1}{\sqrt{N}}=\mathbf{0}^{(N-1)\times1}$  are used.
Since the eigenvectors are orthonormal, we have $V_{\textrm{s}} V_{\textrm{s}}^{T} +\frac{1}{N}\mathbf{1}\mathbf{1}^{T}=\mathbf{I}$.
Left-multiplying (\ref{eq:ekb1}) with $V_{\textrm{s}}^{T}$ gives:\vspace{-5pt}
\begin{align}
& V_{\textrm{s}}^{T} \boldsymbol{e}_{k+\mathfrak{b}-1} = V_{\textrm{s}}^{T} \boldsymbol{\Omega}_{k,k+\mathfrak{b}-1} \nonumber \\
& \qquad +  V_{\textrm{s}}^{T} \mathcal{A}_{k,k+\mathfrak{b}-1} \left( V_{\textrm{s}} V_{\textrm{s}}^{T} +\tfrac{1}{N}\mathbf{1}\mathbf{1}^{T} \right) \boldsymbol{e}_{k-1} \thinspace, \nonumber \\
& = V_{\textrm{s}}^{T} \boldsymbol{\Omega}_{k,k+\mathfrak{b}-1} + V_{\textrm{s}}^{T} \mathcal{A}_{k,k+\mathfrak{b}-1} V_{\textrm{s}} V_{\textrm{s}}^{T} \boldsymbol{e}_{k-1}  \thinspace. \label{eq:system-with-disturbance}
\end{align}
We first investigate the stability of this system without the disturbance term $V_{\textrm{s}}^{T} \boldsymbol{\Omega}_{k,k+\mathfrak{b}-1}$ in (\ref{eq:system-with-disturbance}).
Let $ \| V_{\textrm{s}}^{T} \boldsymbol{e}_{k+\mathfrak{b}-1} \|_2 $ be a candidate Lyapunov function for this system. Therefore, we get:\vspace{-5pt}
\begin{align}
& \| V_{\textrm{s}}^{T} \boldsymbol{e}_{k+\mathfrak{b}-1} \|_2  \leq \| V_{\textrm{s}}^{T} \mathcal{A}_{k,k+\mathfrak{b}-1} V_{\textrm{s}} \|_2 \| V_{\textrm{s}}^{T} \boldsymbol{e}_{k-1} \|_2 \nonumber\\
%& \leq \sqrt{ \lambda_{\max} \left( V_{\textrm{s}}^{T} \mathcal{A}_{k,k+\mathfrak{b}-1}^T \mathcal{A}_{k,k+\mathfrak{b}-1} V_{\textrm{s}}  \right)} \| V_{\textrm{s}}^{T} \boldsymbol{e}_{k-1} \|_2 \thinspace. \nonumber \\
& \leq  \sigma_{\max} ( \mathcal{A}_{k,k+\mathfrak{b}-1} V_{\textrm{s}} )   \| V_{\textrm{s}}^{T} \boldsymbol{e}_{k-1} \|_2 \thinspace, \nonumber \\
& = \sigma_{N-1} (\mathcal{A}_{k,k+\mathfrak{b}-1}) \| V_{\textrm{s}}^{T} \boldsymbol{e}_{k-1} \|_2 \thinspace, \nonumber
\end{align}
where $\sigma_{\max}$ and $\sigma_{N-1}$ denotes the largest and the second largest singular value, respectively. Since $V_{\textrm{s}}^{T}$ is orthonormal (i.e., $V_{\textrm{s}}^{T}V_{\textrm{s}} = \mathbf{I}$) and also orthogonal to $\boldsymbol{1}^T$ (i.e., $V_{\textrm{s}}^{T} \boldsymbol{1} = \boldsymbol{0}$) and the matrix $\mathcal{A}_{k,k+\mathfrak{b}-1}^T \mathcal{A}_{k,k+\mathfrak{b}-1}$ is primitive, we have $\sigma_{\max} ( \mathcal{A}_{k,k+\mathfrak{b}-1} V_{\textrm{s}} ) = \sigma_{N-1} (\mathcal{A}_{k,k+\mathfrak{b}-1})  <1$.
Therefore, the error vector $V_{\textrm{s}}^{T} \boldsymbol{e}_{k}$ is globally exponentially stable in absence of the disturbance term.\\
\phantom{123}Since the matrix $\mathcal{A}_{k,k+\mathfrak{b}-1}$ is irreducible, the matrix $\mathcal{A}_{k,k+\mathfrak{b}(N-1)-1}$ is a positive matrix because the maximum path length between any two agents is less than or equal to $\mathfrak{b}(N-1)$ \citep{Ref:Bandyopadhyay17_TRO}.
Hence the measure of irreducibility of the matrix $\mathcal{A}_{k,k+\mathfrak{b}(N-1)-1}^{T}\mathcal{A}_{k,k+\mathfrak{b}(N-1)-1}$ is lower bounded by $\frac{\gamma-\gamma^{N}}{1-\gamma}$, and we have $\sigma_{N-1}(\mathcal{A}_{k,k+\mathfrak{b}(N-1)-1})\leq\left(1-\frac{4(\gamma-\gamma^{N})}{(1-\gamma)}\sin^{2}\frac{\pi}{2N}\right)^{\frac{1}{2}} < 1$ \citep{Ref:Fiedler72}.
Therefore, $\sigma_{m}$ is given by (\ref{eq:sigma_m_bound}).
Moreover, it follows from Assumption~\ref{assump:bounded_likelihood_func} that
$\left\| \left[ \begin{smallmatrix}
\mathscr{L}_{k}^{1}-\mathscr{L}_{k-1}^{1},
\ldots,
\mathscr{L}_{k}^{N}-\mathscr{L}_{k-1}^{N}
\end{smallmatrix}\right]^T \right\|_2 \leq 2\sqrt{N}\Delta\theta_L$ because $|\mathscr{L}_{k}^{i}-\mathscr{L}_{k-1}^{i}| \leq 2\Delta\theta_L$. Therefore, we have:\vspace{-5pt}
\begin{equation*}
\| V_{\textrm{s}}^{T} \boldsymbol{\Omega}_{k,k+\mathfrak{b}(N-1)-1} \|_2 \leq 2\mathfrak{b}(N-1)\sqrt{N}\Delta\theta_L \thinspace.
\end{equation*}
Hence, in the presence of the disturbance term, we get:\vspace{-5pt}
\begin{align}
& \| V_{\textrm{s}}^{T} \boldsymbol{e}_{k+\mathfrak{b}(N-1)-1} \|_2 \leq \| V_{\textrm{s}}^{T} \boldsymbol{\Omega}_{k,k+\mathfrak{b}(N-1)-1} \|_2  \nonumber \\
& \qquad + \sigma_{N-1} ( \mathcal{A}_{k,k+\mathfrak{b}(N-1)-1}) \| V_{\textrm{s}}^{T} \boldsymbol{e}_{k-1} \|_2 \thinspace, \nonumber \\
& \leq  \sigma_{m}  \| V_{\textrm{s}}^{T} \boldsymbol{e}_{k-1} \|_2  + 2\mathfrak{b}(N-1)\sqrt{N}\Delta\theta_L . \label{eq:conv-with-dist-term}
\end{align}
Using the discrete Gronwall lemma \cite[pp. 9]{Ref:Stuart98} we obtain:\vspace{-5pt}
\begin{align}
\| V_{\textrm{s}}^{T} \boldsymbol{e}_{k} \|_2 & \leq \sigma_{m}^{\left\lfloor \frac{k-1}{\mathfrak{b}(N-1)}\right\rfloor }  \|V_{\textrm{s}}^{T}\boldsymbol{e}_{1}\|_{2}  \nonumber \\
& + \frac{1-\sigma_{m}^{\left\lfloor \frac{k-1}{\mathfrak{b}(N-1)}\right\rfloor }}{1-\sigma_{m}} 2\mathfrak{b}(N-1)\sqrt{N}\Delta\theta_{L} \thinspace. \label{eq:Gronwall}
\end{align}
Moreover, $\|V_{\textrm{s}}^{T}\boldsymbol{e}_{1}(\boldsymbol{x})\|_{2}\leq \sqrt{N}\mathfrak{D}_{1}$, where $\mathfrak{D}_{1}$ is defined above (\ref{eq:kappa1}).
Therefore, it follows that for all $\boldsymbol{x}\in\mathcal{X}$:\vspace{-5pt}
\begin{align}
 & \max_{i\in\mathcal{V}}|\mathscr{U}_{k}^{i}(\boldsymbol{x})-\mathscr{L}_{k}^{\mathrm{KL}}(\boldsymbol{x})|\leq \Xi_k \thinspace,\quad\forall k \in \mathbb{N}\thinspace,\label{eq:proof_step11} \\
 \textrm{where } & \Xi_k = \left(\sqrt{N}\mathfrak{D}_{1}-\tfrac{2\mathfrak{b}(N-1)\sqrt{N}\Delta\theta_{L}}{1-\sigma_{m}}\right)\sigma_{m}^{\left\lfloor \frac{k-1}{\mathfrak{b}(N-1)}\right\rfloor } \nonumber \\
 & \quad +\frac{2\mathfrak{b}(N-1)\sqrt{N}\Delta\theta_{L}}{1-\sigma_{m}}  \thinspace. \label{eq:proof_xi}
\end{align}
Thus, the error between $\mathscr{U}_{k}^{i}$ and $\mathscr{L}_{k}^{\mathrm{KL}}$ is bounded by $\Xi_{k}$, which depends on time instant $k$.

\textbf{Step 2.} We now prove that $\mathcal{T}_{k}^{i}$ (\ref{eq:FODAC-DBFA2})
converges to $\mathcal{L}_{k}^{C}$ (\ref{eq:L-C}).
For all $\boldsymbol{x}\in\mathcal{X}$, (\ref{eq:L-C}) and (\ref{eq:FODAC-DBFA2})
can be re-written as:\vspace{-5pt}
\begin{equation*}
\mathscr{L}_{k}^{C}(\boldsymbol{x})=N\mathscr{L}_{k}^{\mathrm{KL}}(\boldsymbol{x})\thinspace,\quad\mathscr{T}_{k}^{i}(\boldsymbol{x})=N\mathscr{U}_{k}^{i}(\boldsymbol{x})\thinspace,\thinspace\forall i\in\mathcal{V}\thinspace.
\end{equation*}
Therefore, by using (\ref{eq:proof_step11}), we can obtain:\vspace{-5pt}
\begin{equation}
\max_{i\in\mathcal{V}}|\mathscr{T}_{k}^{i}(\boldsymbol{x})-\mathscr{L}_{k}^{C}(\boldsymbol{x})|\leq N\Xi_k \thinspace,\quad \forall k \in \mathbb{N} \thinspace,\label{eq:T-LC-conv-error}\vspace{-5pt}
\end{equation}
\begin{equation*}
\max_{i\in\mathcal{V}}\left|\log\left[\frac{\mathcal{T}_{k}^{i}(\boldsymbol{x})}{\mathcal{T}_{k}^{i}(\boldsymbol{\psi})}\right]-\log\left[\frac{\mathcal{L}_{k}^{C}(\boldsymbol{x})}{\mathcal{L}_{k}^{C}(\boldsymbol{\psi})}\right]\right|\leq N\Xi_k \thinspace,\thinspace \forall k \in \mathbb{N} \thinspace.
\end{equation*}
Using Lemma~\ref{lem:psi-exisits}, we select $\boldsymbol{\psi}\in\mathcal{X}$ such that $\mathcal{T}_{k}^{i}(\boldsymbol{\psi})=\mathcal{L}_{k}^{C}(\boldsymbol{\psi})$. Therefore,\vspace{-5pt}
\begin{align*}
\max_{i\in\mathcal{V}}\left|\log\left[\frac{\mathcal{T}_{k}^{i}(\boldsymbol{x})}{\mathcal{L}_{k}^{C}(\boldsymbol{x})}\right]\right| & \leq N\Xi_k \thinspace,\thinspace \forall k \in \mathbb{N} \thinspace,\\
e^{-N\Xi_k }\leq\max_{i\in\mathcal{V}}\left(\frac{\mathcal{T}_{k}^{i}(\boldsymbol{x})}{\mathcal{L}_{k}^{C}(\boldsymbol{x})}\right) & \leq e^{N\Xi_k }\thinspace,\thinspace \forall k \in \mathbb{N} \thinspace.
\end{align*}
\begin{equation*}
\max_{i\in\mathcal{V}}\left|\mathcal{T}_{k}^{i}(\boldsymbol{x})-\mathcal{L}_{k}^{C}(\boldsymbol{x})\right|\leq \mathcal{L}_{k}^{C}(\boldsymbol{x})\left(e^{N\Xi_{k}}-1\right)\thinspace,\thinspace\forall k\in\mathbb{N}\thinspace.
\end{equation*}
Since $\boldsymbol{x}\in\mathcal{X}$ can be any point, therefore:\vspace{-5pt}
\begin{align}
 & \max_{i\in\mathcal{V}}D_{L_{1}}\left(\mathcal{T}_{k}^{i},\mathcal{L}_{k}^{C}\right)=\max_{i\in\mathcal{V}}\int_{\mathcal{X}}\left|\mathcal{T}_{k}^{i}-\mathcal{L}_{k}^{C}\right|\:d\mu(\boldsymbol{x})\nonumber \\
 & \leq \left(e^{N\Xi_{k}}-1\right) \int_{\mathcal{X}}\mathcal{L}_{k}^{C}\:d\mu(\boldsymbol{x})= \left(e^{N\Xi_{k}}-1\right) \thinspace,\thinspace\forall k\in\mathbb{N} \thinspace. \nonumber % \label{eq:proof_step9}
\end{align}
Hence the convergence error is bounded by $\left(e^{N\Xi_{k}}-1\right)$.\\
\phantom{123}It follows from (\ref{eq:T_error-ss})--(\ref{eq:T_error}) that $\left(e^{N\Xi_{k}}-1\right)\leq(1+\eta)\delta$ for all $k\geq\kappa$ and $\lim_{k\rightarrow\infty}\left(e^{N\Xi_{k}}-1\right)\leq\delta$.
The time step size $\Delta$ (\ref{eq:Delta_max}) is found using the steady-state error term:\vspace{-5pt}
\begin{align}
\exp{\left(N\frac{2\mathfrak{b}(N-1)\sqrt{N}\Delta\theta_{L}}{1-\sigma_{m}}\right)} - 1&=\delta \thinspace . \label{eq:steady-state-error-term}
\end{align}
$\delta_{\mathrm{min}}$ (\ref{eq:delta_min}) is obtained by substituting $\Delta_{\mathrm{min}}$ into (\ref{eq:steady-state-error-term}).
If $\sqrt{N}\mathfrak{D}_{1} \leq \frac{2\mathfrak{b}(N-1)\sqrt{N} \Delta \theta_{L}}{1-\sigma_{m}}$, then $\left(e^{N\Xi_{k}}-1\right)\leq(1+\eta)\delta$ for all $k \in \mathbb{N}$. Therefore, if $\mathfrak{D}_{1}\leq \frac{\log\left(\delta+1\right)}{N^{\frac{3}{2}}}$, then $\kappa = 1$.
Otherwise, for $\mathfrak{D}_{1} > \frac{\log\left(\delta+1\right)}{N^{\frac{3}{2}}}$, $\kappa$  (\ref{eq:kappa1}) is computed using $\left(e^{N\Xi_{k}}-1\right)\leq(1+\eta)\delta$.
The constraint on TV error follows from Lemma~\ref{lem:convegence-TV}. Our exponential stability proof is substantially different from the asymptotic-convergence proof in~\citep{Ref:Zhu10}. \hfill $\blacksquare$

%Theorem \ref{thm:conv-likelihood} explicitly bounds the time step
%$\Delta$ of the distributed estimation algorithm with the time-scale
%of the target dynamics using Assumption \ref{thm:conv-likelihood}.
%Moreover, after $\kappa$ time instants, each agent's estimated likelihood
%function $\mathcal{T}_{k}^{i}$ converges to an error ball centered
%on the normalized joint likelihood function $\mathcal{L}_{k}^{C}$
%used in the multi-sensor Bayesian filtering algorithm (\ref{eq:update_stage_multisensor}).
%Therefore, the $i^{\textrm{th}}$ agent uses its estimated likelihood
%function $\mathcal{T}_{k}^{i}$ to compute the posterior pdf $\mathcal{W}_{k}^{i}$
%during the update step of the Bayesian filtering algorithm:\vspace{-5pt}

\begin{remark} A key advantage of the DBF algorithm is that it does
not require all the sensors to observe the target. If an agent does
not observe the target, then it sets its normalized likelihood function
as the uniform distribution, i.e., $\mathcal{L}_{k}^{i}(\boldsymbol{x})=1$.
%\frac{1}{\int_{\mathcal{X}}\thinspace d\mu(\boldsymbol{x})}$.
Then this agent's likelihood function does not influence the joint
likelihood function and the estimated pdfs because of the geometric
nature of the fusion rule. Moreover, the DBF algorithm avoids double counting because the summation of weights from all paths is a constant due to the weights in the adjacency matrix $\mathcal{A}_k$.
Theorem \ref{thm:conv-likelihood} explicitly bounds the time step size
$\Delta$ of the distributed estimation algorithm with the time-scale
of the target dynamics.
But the effectiveness of the DBF algorithm is
predicated on Assumption \ref{assump:bounded_likelihood_func}. Moreover,
the upper bound on the time step size $\Delta_{\max}$ (\ref{eq:Delta_max})
decreases with increasing number of agents $N$. \end{remark}
%In the following corollary, we provide sharper bounds for $\kappa$ (\ref{eq:kappa1}) and $\Delta_{\max}$ (\ref{eq:Delta_max}) for a special case of the communication network topology.
%%In the special case where the communication network topology is strongly connected and time-invariant, the following corollary gives sharper bounds for $\kappa$ (\ref{eq:kappa1}) and $\Delta_{\max}$ (\ref{eq:Delta_max}).

The following corollary provides sharper bounds for the special case of a static, strongly-connected communication network topology.
\begin{corollary} \label{cor:conv-likelihood-A}
If the communication network topology is time-invariant and strongly-connected, the time step size $\Delta$ (\ref{eq:Delta_max}), $\delta_\mathrm{min}$ (\ref{eq:delta_min}), and $\kappa$ (\ref{eq:kappa1}) in Theorem~\ref{thm:conv-likelihood} are given by:\vspace{-5pt}
\begin{align}
& \Delta = \frac{\left(1-\sigma_{N-1}(\mathcal{A})\right)\log\left(\delta+1\right)}{2N\sqrt{N}\theta_{L}} \thinspace, \label{eq:Delta_max_A} \\
& \delta_{\mathrm{min}} = \exp{\left( \frac{\Delta_{\mathrm{min}} 2N\sqrt{N}\theta_{L}}{1-\sigma_{N-1}(\mathcal{A})} \right) } - 1 \thinspace, \label{eq:delta_min_A} \\
&\kappa= \left\lceil \tfrac{1}{\log\sigma_{N-1}(\mathcal{A})}\log\left(\frac{\log\left(\frac{(1+\eta)\delta+1}{\delta+1}\right)}{\log\left(\frac{e^{N^{\frac{3}{2}}\mathfrak{D}_{1}}}{\delta+1}\right)}\right)\right\rceil  +1,   \label{eq:kappa_new_A}
\end{align}
where $\mathcal{A}$ is the time-invariant adjacency matrix.
\end{corollary}
\emph{Proof:} In this case, (\ref{eq:conv-with-dist-term}) is written as:\vspace{-5pt}
\begin{align*}
& \|V_{\textrm{s}}^{T}\boldsymbol{e}_{k}\|_{2} \leq \sigma_{N-1}(\mathcal{A})\|V_{\textrm{s}}^{T}\boldsymbol{e}_{k-1}\|_{2}+2\sqrt{N}\Delta\theta_{L}  \thinspace.
\end{align*}
Using the discrete Gronwall lemma \cite[pp. 9]{Ref:Stuart98} we obtain:\vspace{-5pt}
\begin{align}
\|V_{\textrm{s}}^{T}\boldsymbol{e}_{k}\|_{2} & \leq\left(\sigma_{N-1}(\mathcal{A})\right)^{k-1}\|V_{\textrm{s}}^{T}\boldsymbol{e}_{1}\|_{2} \nonumber \\
& +\frac{1-\left(\sigma_{N-1}(\mathcal{A})\right)^{k-1}}{1-\sigma_{N-1}(\mathcal{A})}2\sqrt{N}\Delta\theta_{L} \thinspace.
\end{align}
Hence, we get $\max_{i\in\mathcal{V}}|\mathscr{U}_{k}^{i}(\boldsymbol{x})-\mathscr{L}_{k}^{\mathrm{KL}}(\boldsymbol{x})|\leq\Xi_k $ for all $k \in \mathbb{N}$, where\vspace{-5pt}
\begin{align}
\Xi_k &= \left(\sigma_{N-1}(\mathcal{A})\right)^{k-1}\sqrt{N}\mathfrak{D}_{1}+\tfrac{1-\left(\sigma_{N-1}(\mathcal{A})\right)^{k-1}}{1-\sigma_{N-1}(\mathcal{A})}2\sqrt{N}\Delta\theta_{L}  \thinspace. \nonumber
\end{align}
We get $\Delta$ (\ref{eq:Delta_max_A}) and $\delta_{\mathrm{min}}$ (\ref{eq:delta_min_A}) from $\lim_{k\rightarrow\infty}\left(e^{N\Xi_{k}}-1\right)\leq\delta$ and $\kappa$ (\ref{eq:kappa_new_A}) from $\left(e^{N\Xi_{k}}-1\right)\leq(1+\eta)\delta$ for all $k\geq\kappa$.  \hfill $\blacksquare$

Note that $\Delta$ (\ref{eq:Delta_max_A}), $\delta_{\mathrm{min}}$ (\ref{eq:delta_min_A}), and $\kappa$ (\ref{eq:kappa_new_A}) in Corollary~\ref{cor:conv-likelihood-A}
can be obtained from $\Delta$ (\ref{eq:Delta_max}), $\delta_\mathrm{min}$ (\ref{eq:delta_min}), and $\kappa$ (\ref{eq:kappa1}) in Theorem~\ref{thm:conv-likelihood} by replacing $\mathfrak{b}(N-1)$ with $1$.

\subsection{Robustness Analysis \label{sub:Robustness}}
The agents need to communicate their pdfs $\mathcal{U}_{k-1}^j$ with their neighbors (see line 5 in Algorithm~\ref{alg:DBF}).
\begin{remark} (Communication of pdfs) The information theoretic approach for communicating pdfs is studied in \citep{Ref:Kramer07}.
If particle filters are used to implement the Bayesian filter and combine the pdfs \citep{Ref:Arulampalam02}, then the resampled particles represent the agent's estimated pdf.
Hence communicating pdfs is equivalent
to transmitting these resampled particles. Another approach involves
approximating the pdf by a weighted sum of Gaussian pdfs \cite[pp. 213]{Ref:Anderson05}
and then transmitting this approximate distribution. Several techniques
for estimating the Gaussian parameters are discussed in the Gaussian
mixture model literature \citep{Ref:Kotecha03,Ref:McLachlan88,Ref:Reynolds08}. \end{remark}

Let the pdf $\hat{\mathcal{U}}_{k}^{i}\in\Phi(\mathcal{X})$ denote the pdf $\mathcal{U}_{k}^{i}$ that is corrupted with communication errors. Similarly, let the pdf $\hat{\mathcal{L}}_{k}^{i}\in\Phi(\mathcal{X})$ represent the normalized likelihood function $\mathcal{L}_{k}^{i}$ that is corrupted with modeling errors. We first state the assumptions on these errors and then state the main result of this section.

\begin{assumption} \label{assump:bounded_comm_error} There exists time-invariant constants
$\varepsilon_{U}\geq0$ and
$\varepsilon_{L}\geq0$ such
that for all agents $i\in\mathcal{V}$:\vspace{-5pt}
\begin{align}
e^{-\varepsilon_{U}} & \leq\frac{\hat{\mathcal{U}}_{k}^{i}(\boldsymbol{x})}{\mathcal{U}_{k}^{i}(\boldsymbol{x})}\leq e^{\varepsilon_{U}}\thinspace,\qquad\forall\boldsymbol{x}\in\mathcal{X},\thinspace\forall k\in\mathbb{N}\thinspace,\label{eq:bounded_comm_error}\\
e^{-\varepsilon_{L}} & \leq\frac{\hat{\mathcal{L}}_{k}^{i}(\boldsymbol{x})}{\mathcal{L}_{k}^{i}(\boldsymbol{x})}\leq e^{\varepsilon_{L}}\thinspace,\qquad\forall\boldsymbol{x}\in\mathcal{X},\thinspace\forall k\in\mathbb{N}\thinspace.\label{eq:bounded_rep_error}
\end{align}
Therefore, $|\mathscr{U}_{k}^{i} - \hat{\mathscr{U}}_{k}^{i}| \leq 2\varepsilon_{U} $ and $|\mathscr{L}_{k}^{i} - \hat{\mathscr{L}}_{k}^{i}| \leq 2\varepsilon_{L} $, where $\hat{\mathscr{U}}_{k}^{i}(\boldsymbol{x}) =\log\left[\frac{\hat{\mathcal{U}}_{k}^{i}(\boldsymbol{x})}{\hat{\mathcal{U}}_{k}^{i}(\boldsymbol{\psi})}\right]$ and $\hat{\mathscr{L}}_{k}^{i}(\boldsymbol{x}) =\log\left[\frac{\hat{\mathcal{L}}_{k}^{i}(\boldsymbol{x})}{\hat{\mathcal{L}}_{k}^{i}(\boldsymbol{\psi})}\right]$.
\end{assumption}

\begin{corollary} \label{cor:Robustness} Under Assumptions \ref{assump:comm-topology}--\ref{assump:bounded_comm_error}, the time step size $\Delta$ (\ref{eq:Delta_max}) and $\delta_\mathrm{min}$ (\ref{eq:delta_min}) in Theorem~\ref{thm:conv-likelihood} is given by:\vspace{-5pt}
\begin{align}
& \Delta = \frac{\left(1-\sigma_{m}\right)\log\left(\delta+1\right)}{2\mathfrak{b}N(N-1)\sqrt{N}\theta_{L}}-\frac{2\varepsilon_{L}+\varepsilon_{U}}{\theta_{L}} \thinspace, \label{eq:Delta_max_comm} \\
& \delta_{\mathrm{min}} = e^{\left(\Delta_{\mathrm{min}} + \frac{2\varepsilon_{L}+\varepsilon_{U}}{\theta_{L}} \right) \left( \frac{ 2\mathfrak{b}N(N-1)\sqrt{N}\theta_{L}}{1-\sigma_{m}} \right) }  - 1 \thinspace, \label{eq:delta_min_comm}
\end{align}
where $\varepsilon_{U}$ and $\varepsilon_{L}$ are defined in Assumption~\ref{assump:bounded_comm_error}.
\end{corollary}

\emph{Proof:}
Equation (\ref{eq:FODAC-DBFA1}) can be written as:\vspace{-5pt}
\begin{align}
\mathscr{U}_{k}^{i} & \!=\! \begin{cases}
\hat{\mathscr{L}}_{1}^{i} & \textrm{if } k=1\\
\sum_{j=1,j\not=i}^{N} \mathcal{A}_{k}[i,j]\hat{\mathscr{U}}_{k-1}^{j}  &  \\
+\! \mathcal{A}_{k}[i,i]\mathscr{U}_{k-1}^{i} +\! \hat{\mathscr{L}}_{k}^{i} \!-\! \hat{\mathscr{L}}_{k-1}^{i} & \textrm{if } k\geq2
\end{cases}, \label{eq:proof_step1_comm}
\end{align}
Substituting the bounds from Assumption~\ref{assump:bounded_comm_error} gives:\vspace{-5pt}
\begin{align*}
& | \mathscr{U}_{1}^{i} - \mathscr{L}_{1}^{i} | \leq 2\varepsilon_{L} \thinspace,  \\
& | \mathscr{U}_{k}^{i} - \sum_{j=1}^{N}\! \mathcal{A}_{k}[i,j]\mathscr{U}_{k-1}^{j} \!-\! \mathscr{L}_{k}^{i} \!+\! \mathscr{L}_{k-1}^{i} | \leq 2\varepsilon_{U} + 4\varepsilon_{L} \thinspace.
\end{align*}
The evolution of the error vector $\boldsymbol{e}_k$ is now given by:\vspace{-5pt}
\begin{align}
\boldsymbol{e}_k & \!=\! \mathcal{A}_{k} \boldsymbol{e}_{k-1} + \hat{\boldsymbol{\Omega}}_{k,k} \thinspace, \qquad \forall k \geq 2 \thinspace,  \\
\textrm{where } \| \hat{\boldsymbol{\Omega}}_{k,k} \|_2 & \leq \| \boldsymbol{\Omega}_{k,k} \|_2 + 2\sqrt{N}(\varepsilon_{U} + 2\varepsilon_{L}) \thinspace. \nonumber
\end{align}
Similar to the proof of Theorem~\ref{thm:conv-likelihood}, we get:\vspace{-5pt}
\begin{align}
& \| V_{\textrm{s}}^{T} \boldsymbol{e}_{k} \|_2 \leq \sigma_{m}^{\left\lfloor \frac{k-1}{\mathfrak{b}(N-1)}\right\rfloor }  \|V_{\textrm{s}}^{T}\boldsymbol{e}_{1}\|_{2}  \nonumber \\
& + \tfrac{1-\sigma_{m}^{\left\lfloor \frac{k-1}{\mathfrak{b}(N-1)}\right\rfloor }}{1-\sigma_{m}} 2\mathfrak{b}(N-1)\sqrt{N} (\Delta\theta_{L} + 2\varepsilon_{L} + \varepsilon_{U}) \thinspace. \nonumber
\end{align}
Hence, we get $\max_{i\in\mathcal{V}}|\mathscr{U}_{k}^{i}(\boldsymbol{x})-\mathscr{L}_{k}^{\mathrm{KL}}(\boldsymbol{x})|\leq\Xi_k$ for all $k \in \mathbb{N}$, where\vspace{-5pt}
\begin{align}
& \Xi_k = \sigma_{m}^{\left\lfloor \frac{k-1}{\mathfrak{b}(N-1)}\right\rfloor }\sqrt{N}\mathfrak{D}_{1} \nonumber \\
 & +\tfrac{1-\sigma_{m}^{\left\lfloor \frac{k-1}{\mathfrak{b}(N-1)}\right\rfloor }}{1-\sigma_{m}}2\mathfrak{b}(N-1)\sqrt{N} (\Delta\theta_{L} + 2\varepsilon_{L} + \varepsilon_{U}) \thinspace. \nonumber
\end{align}
We get $\Delta$ (\ref{eq:Delta_max_comm}) and $\delta_\mathrm{min}$ (\ref{eq:delta_min_comm}) from $\lim_{k\rightarrow\infty}\left(e^{N\Xi_{k}}-1\right)\leq\delta$.
We get the same $\kappa$ (\ref{eq:kappa1}) for this case.   \hfill $\blacksquare$

It follows from Corollary~\ref{cor:Robustness} that in order to generate satisfactory estimates using the DBF algorithm, the bounds $\varepsilon_{U},\thinspace \varepsilon_{L}$ should be substantially smaller than $\delta$.

\subsection{Distributed Kalman Information Filter \label{sec:Special-Case-Gaussian}}
The DBF algorithm is applied to linear target dynamics and measurement models with additive Gaussian noise:\vspace{-5pt}
\begin{align}
\boldsymbol{x}_{k+1} &=\boldsymbol{F}_{k}\boldsymbol{x}_{k}+\boldsymbol{w}_{k}\thinspace, & & \quad \forall k \in \mathbb{N} \thinspace, \label{eq:sys_mod_linear} \\
\boldsymbol{y}_{k}^{i} &=\boldsymbol{H}_{k}^{i}\boldsymbol{x}_{k}+\boldsymbol{v}_{k}^{i}\thinspace,  & & \quad \forall k \in \mathbb{N} \thinspace, \thinspace \forall i\in\mathcal{V}\thinspace,\label{eq:mes_mod_linear}
\end{align}
where the process noise $\boldsymbol{w}_{k} = \mathcal{N}( \boldsymbol{0}, \boldsymbol{Q}_{k})$ and the measurement noise $\boldsymbol{v}_{k}^{i} = \mathcal{N}( \boldsymbol{0}, \boldsymbol{R}_{k}^{i})$ are zero mean multivariate normal distributions.
Therefore, we adopt the information filter-based representation \citep{Ref:Mutambara98,Ref:Fourati15}. The pseudo-code of the distributed Kalman information filtering algorithm for linear-Gaussian models is given in Algorithm~\ref{alg:DBF-LGM}.
The prior pdf $\mathcal{S}_{k}^{i}= \mathcal{N}(\hat{\boldsymbol{x}}_{k|k-1}^{i},\boldsymbol{P}_{k|k-1}^{i})$, the posterior pdf $\mathcal{W}_{k}^{i} = \mathcal{N}(\hat{\boldsymbol{x}}_{k|k}^{i},\boldsymbol{P}_{k|k}^{i})$, and the estimated pdfs  $\mathcal{U}_{k}^{i} = \mathcal{N} \left( (\boldsymbol{U}_{k}^{i})^{-1} \boldsymbol{u}_{k}^{i} , (\boldsymbol{U}_{k}^{i})^{-1} \right)$, $\mathcal{T}_{k}^{i} = \mathcal{N} \left( (\boldsymbol{T}_{k}^{i})^{-1} \boldsymbol{t}_{k}^{i} , (\boldsymbol{T}_{k}^{i})^{-1} \right)$ are also multivariate normal distributions.
\begin{algorithm}[t]
\caption{Distributed Kalman Information Filtering \label{alg:DBF-LGM}}
\centering{}%
\resizebox{3.4in}{!}{
\begin{tabular}{ll}
\hline
1. & ($i^{\textrm{th}}$ agent's steps at $k^{\textrm{th}}$ time instant)\tabularnewline
2. & Compute the prior pdf $\mathcal{S}_{k}^{i} = \mathcal{N}(\hat{\boldsymbol{x}}_{k|k-1}^{i},\boldsymbol{P}_{k|k-1}^{i})$: \vspace{-10pt} \tabularnewline
 & \vbox{ \begin{align}
 \hat{\boldsymbol{z}}_{k-1|k-1}^{i} & =(\boldsymbol{P}_{k-1|k-1}^{i})^{-1}\hat{\boldsymbol{x}}_{k-1|k-1}^{i}\thinspace, \nonumber \\
 \boldsymbol{Z}_{k-1|k-1}^{i} &=(\boldsymbol{P}_{k-1|k-1}^{i})^{-1}\thinspace,\nonumber \\
\boldsymbol{M}_{k-1}^{i}&=(\boldsymbol{F}_{k-1}^{-1})^{T}\boldsymbol{Z}_{k-1|k-1}^{i}\boldsymbol{F}_{k-1}^{-1}\thinspace, \nonumber \\
\boldsymbol{Z}_{k|k-1}^{i}&= \left(\mathbf{I}-\boldsymbol{M}_{k-1}^{i}\left(\boldsymbol{M}_{k-1}^{i}+ \boldsymbol{Q}_{k-1}^{-1}\right)^{-1}\right)\boldsymbol{M}_{k-1}^{i}\thinspace,  \nonumber \\
\hat{\boldsymbol{z}}_{k|k-1}^{i}&=\left(\mathbf{I}-\boldsymbol{M}_{k-1}^{i}\left(\boldsymbol{M}_{k-1}^{i}+ \boldsymbol{Q}_{k-1}^{-1}\right)^{-1}\right)(\boldsymbol{F}_{k-1}^{-1})^{T}\hat{\boldsymbol{z}}_{k-1|k-1}^{i}\thinspace, \nonumber \\
\boldsymbol{P}_{k|k-1}^{i} &= ( \boldsymbol{Z}_{k|k-1}^{i} )^{-1} \thinspace,  \qquad
\hat{\boldsymbol{x}}_{k|k-1}^{i} = \boldsymbol{P}_{k|k-1}^{i}\hat{\boldsymbol{z}}_{k|k-1}^{i} \thinspace. \nonumber
\end{align}} \vspace{-10pt} \tabularnewline
3. & Obtain local measurement $\boldsymbol{y}_{k}^{i}$.\tabularnewline
4. & Receive pdfs $\mathcal{U}_{k-1}^{j}$ from agents $j\in\mathcal{J}_{k}^{i}$.\tabularnewline
5. & Compute the pdfs $\mathcal{U}_{k}^{i}$ and $\mathcal{T}_{k}^{i}$ as follows:\vspace{-10pt} \tabularnewline
 & \vbox{\begin{align}
\boldsymbol{i}_{k}^{i} & =(\boldsymbol{H}_{k}^{i})^{T}(\boldsymbol{R}_{k}^{i})^{-1}\boldsymbol{y}_{k}^{i}\thinspace, \nonumber\\ %\label{eq:z_k}\\
\boldsymbol{I}_{k}^{i} & =(\boldsymbol{H}_{k}^{i})^{T}(\boldsymbol{R}_{k}^{i})^{-1}\boldsymbol{H}_{k}^{i}\thinspace, \nonumber \\ %\label{eq:Z_k} \\
\boldsymbol{u}_{k}^{i} &=\begin{cases}
\boldsymbol{i}_{1}^{i} & \textrm{ if }k=1\\
\boldsymbol{i}_{k}^{i}-\boldsymbol{i}_{k-1}^{i}+\sum_{j\in\mathcal{J}_{k}^{i}}\mathcal{A}_{k}[i,j]\boldsymbol{u}_{k-1}^{j}\thinspace, & \textrm{ if }k\geq2
\end{cases}\thinspace, \nonumber \\ %\label{eq:s_k}\\
\boldsymbol{U}_{k}^{i} &= \begin{cases}
\boldsymbol{I}_{1}^{i} & \textrm{ if }k=1\\
\boldsymbol{I}_{k}^{i}-\boldsymbol{I}_{k-1}^{i}+\sum_{j\in\mathcal{J}_{k}^{i}}\mathcal{A}_{k}[i,j]\boldsymbol{U}_{k-1}^{j} \thinspace, & \textrm{ if }k\geq2
\end{cases}\thinspace, \nonumber \\
\boldsymbol{t}_{k}^{i} & =N\boldsymbol{u}_{k}^{i}\thinspace, \qquad \boldsymbol{T}_{k}^{i} =N\boldsymbol{U}_{k}^{i}\thinspace, \nonumber
%\mathcal{U}_{k}^{i} &= \mathcal{N} \left( (\boldsymbol{U}_{k}^{i})^{-1} \boldsymbol{u}_{k}^{i} , (\boldsymbol{U}_{k}^{i})^{-1} \right) \thinspace, \\
%\mathcal{T}_{k}^{i} &= \mathcal{N} \left( (\boldsymbol{T}_{k}^{i})^{-1} \boldsymbol{t}_{k}^{i} , (\boldsymbol{T}_{k}^{i})^{-1} \right) \thinspace.
\end{align}} \vspace{-10pt} \tabularnewline
%4. & Compute $\boldsymbol{I}_{k}^{i}$ and $\boldsymbol{i}_{k}^{i}$ using
%(\ref{eq:z_k})--(\ref{eq:Z_k}) \tabularnewline
%6. & Compute $\boldsymbol{U}_{k}^{i}$, $\boldsymbol{u}_{k}^{i}$, $\boldsymbol{T}_{k}^{i}$,
%and $\boldsymbol{t}_{k}^{i}$ using (\ref{eq:s_k})--(\ref{eq:t_k})\tabularnewline
%7. & Compute posterior $\boldsymbol{Z}_{k|k}^{i}$ and $\hat{\boldsymbol{z}}_{k|k}^{i}$ using (\ref{eq:A_k_k})\tabularnewline
6. & Compute the posterior pdf $\mathcal{W}_{k}^{i}=\mathcal{N}(\hat{\boldsymbol{x}}_{k|k}^{i},\boldsymbol{P}_{k|k}^{i})$: \vspace{-10pt} \tabularnewline
 & \vbox{\begin{align}
\hat{\boldsymbol{z}}_{k|k}^{i} & =\hat{\boldsymbol{z}}_{k|k-1}^{i}+\boldsymbol{t}_{k}^{j}\thinspace, \qquad
\boldsymbol{Z}_{k|k}^{i} =\boldsymbol{Z}_{k|k-1}^{i}+\boldsymbol{T}_{k}^{j}\thinspace, \nonumber \\
\boldsymbol{P}_{k|k}^{i} &=(\boldsymbol{Z}_{k|k}^{i})^{-1} \thinspace, \qquad
\hat{\boldsymbol{x}}_{k|k}^{i} =\boldsymbol{P}_{k|k}^{i}\hat{\boldsymbol{z}}_{k|k}^{i} \thinspace. \nonumber
\end{align}} \vspace{-10pt} \tabularnewline
\hline
\end{tabular}
}\vspace{-5pt}
\end{algorithm}
\subsection{Multiple Consensus Loops within Each Time}
In this section, we show that the proposed DBF algorithm can be easily extended to recursively combine local likelihood functions using multiple consensus loops within each time instant so that each agent's estimated likelihood function converges to the joint likelihood function $\mathcal{L}_{k}^{C}$ (\ref{eq:L-C}). Then, the resultant DBF algorithm is equivalent to the Bayesian consensus algorithms in \citep{Ref:Hlinka12,Ref:Hlinka14}. Note that multiple consensus loops within each time step significantly reduces the practicality of such algorithms.
Let the pdfs $\mathcal{U}_{k,\nu}^{i}\in\Phi(\mathcal{X})$ and $\mathcal{T}_{k,\nu}^{i}\in\Phi(\mathcal{X})$
denote to the local pdfs of the $i^{\textrm{th}}$ agent during the
$\nu^{\textrm{th}}$ consensus loop at the $k^{\textrm{th}}$ time
instant. Since the pdf $\mathcal{L}_{k}^{i}$ is not updated during
the $k^{\textrm{th}}$ time instant, we define the pdfs $\mathcal{L}_{k,\nu}^{i}=\mathcal{L}_{k}^{i}$
for all $\nu\in\mathbb{N}$.
During the $\nu^{\textrm{th}}$ consensus loop, each agent updates its local pdfs
$\mathcal{U}_{k,\nu}^{i}$ and $\mathcal{T}_{k,\nu}^{i}$ using the following fusion rule:\vspace{-5pt}
\begin{align}
 & \mathcal{U}_{k,\nu}^{i}=\begin{cases}
\mathcal{L}_{k,1}^{i} & \textrm{ if }\nu=1\\
\frac{ \prod\limits _{j\in\mathcal{J}_{k}^{i}}(\mathcal{U}_{k,\nu-1}^{j})^{\mathcal{A}_{k}[i,j]} }{\int_{\mathcal{X}} \prod\limits _{j\in\mathcal{J}_{k}^{i}}(\mathcal{U}_{k,\nu-1}^{j})^{\mathcal{A}_{k}[i,j]} \thinspace d\mu(\boldsymbol{x})} & \textrm{ if }\nu\geq2
\end{cases}\thinspace, \\
 & \mathcal{T}_{k}^{i}=\frac{(\mathcal{U}_{k,\nu}^{i})^{N}}{\int_{\mathcal{X}}(\mathcal{U}_{k,\nu}^{i})^{N}d\mu(\boldsymbol{x})}\thinspace.
\end{align}
\begin{theorem} \citep{Ref:Bandyopadhyay14_ACC_BCF,Ref:Bandyopadhyay_BCF_arxiv} Assuming $\mathcal{G}_{k}$ is strongly connected, each agent's pdf $\mathcal{T}_{k,\nu}^{i}$ globally exponentially converges pointwise to $\mathcal{L}_{k}^{C}$ (\ref{eq:L-C}).
%the normalized joint likelihood function $\mathcal{L}_{k}^{C}$ (\ref{eq:L-C}).
After $n_\mathrm{loop}$ consensus loops, the $\ell_2$ norm of the error vector $\boldsymbol{e}_{k,\nu}:=\left[ D_{L_{1}}(\mathcal{T}_{k,\nu}^{1}, \mathcal{L}_{k}^{C}), \ldots, D_{L_{1}}(\mathcal{T}_{k,\nu}^{N}, \mathcal{L}_{k}^{C})  \right]^T$ is bounded by $ \| \boldsymbol{e}_{k,n_\mathrm{loop}} \|_2  \leq ( \sigma_{N-1}(\mathcal{A}_k) )^{(n_\mathrm{loop}-1)} 2\sqrt{N} $. \end{theorem}
% at a rate faster or equal
%to the second-largest singular value of the doubly stochastic matrix
%$\mathcal{A}_{k}$, where the communication network topology is strongly
%connected.
The proof follows from Theorem 2 and 4 in \citep{Ref:Bandyopadhyay14_ACC_BCF}.
%of global exponentially convergence and a relationship between number of consensus loops and convergence error are given in \citep{Ref:Bandyopadhyay14_ACC_BCF,Ref:Bandyopadhyay_BCF_arxiv}.
Thus, the distributed estimation
algorithm in \citep{Ref:Hlinka12,Ref:Hlinka14} is a special case of
our DBF algorithm.

\section{Numerical Simulations \label{sec:Numerical-Simulations}}
In this section, we demonstrate the properties of the DBF algorithm using a benchmark example in Section~\ref{sub:Sim-Benchmark-Problem} and a complex multi-agent estimation and control task in Section~\ref{sub:Sim-Multiagent-Formation}.
%In subsection~\ref{sub:Sim-Benchmark-Problem}, we compare the performance of the DBF algorithms with centralized multi-sensor Bayesian filtering algorithms using the benchmark problem studied in \citep{Ref:Battistelli15,Ref:Battistelli14,Ref:BarShalom04}.
%In subsection~\ref{sub:Sim-Multiagent-Formation}, we demonstrate the effectiveness of the DBF algorithm using a complex formation control problem.

\subsection{Benchmark Example \label{sub:Sim-Benchmark-Problem}}
\begin{figure}[!h]
\begin{centering}
\includegraphics[width=2.5in]{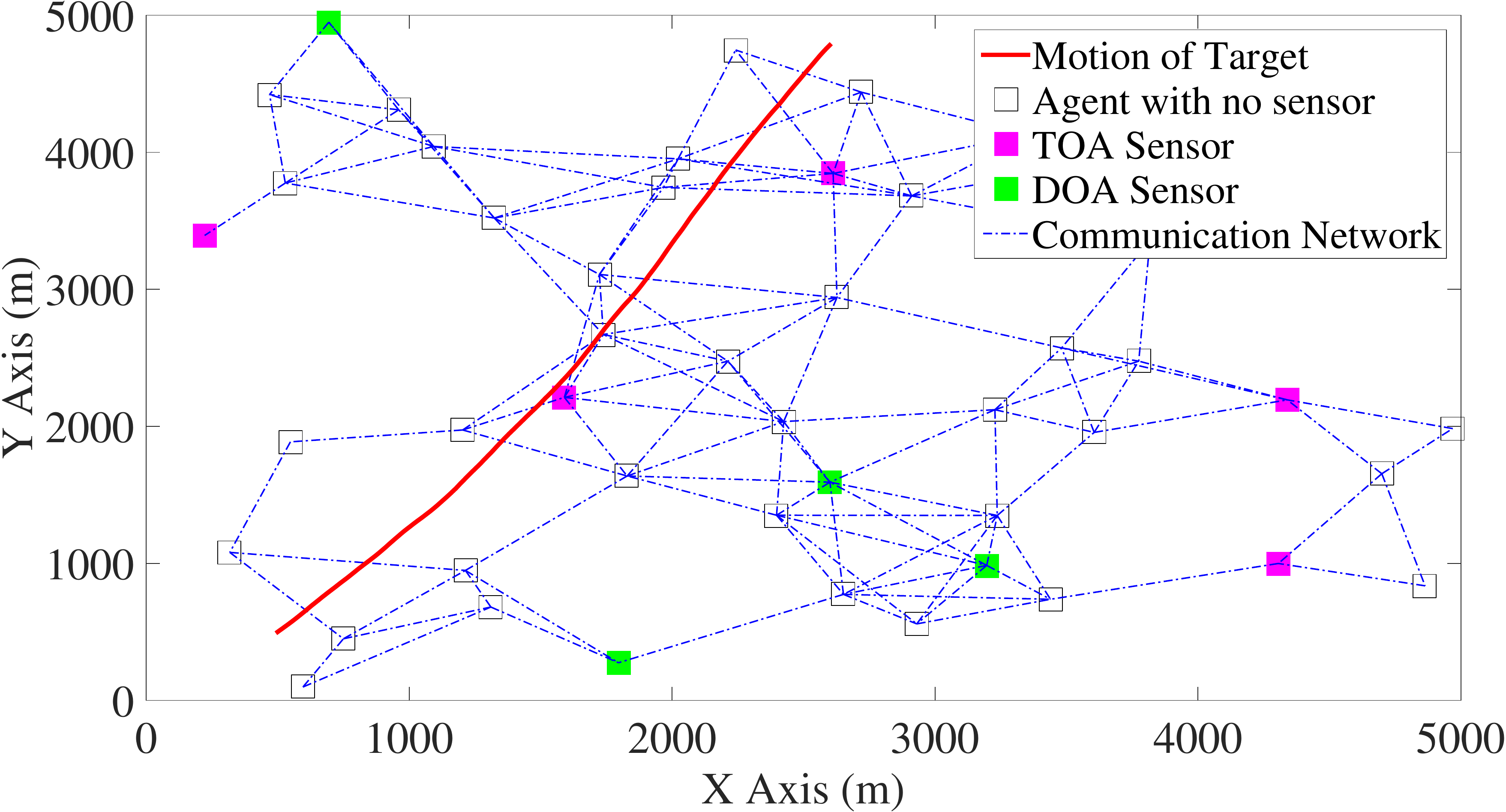}
\par\end{centering}
\caption{The motion of the target, the position of sensing agents (5 TOA sensors,
5 DOA sensors, and 40 agents with no sensors), and their communication
network topology. \label{fig:Comm-topology}}
\end{figure}

In this subssection, we compare the performance of the DBF algorithms with the centralized multi-sensor Bayesian filtering algorithms using the benchmark example studied in \citep{Ref:Battistelli15,Ref:Battistelli14,Ref:BarShalom04}.
%The target's motion is shown in Fig.~\ref{fig:Comm-topology}.
The target dynamics is modeled by a linear
model: \vspace{-10pt}
\begin{align*}
\boldsymbol{x}_{k+1} & =\left[\begin{smallmatrix}1 & \Delta & 0 & 0\\
0 & 1 & 0 & 0\\
0 & 0 & 1 & \Delta\\
0 & 0 & 0 & 1
\end{smallmatrix}\right]\boldsymbol{x}_{k}+\boldsymbol{w}_{k}\thinspace, \textrm{where } \boldsymbol{Q} =\left[\begin{smallmatrix}\frac{\Delta^{3}}{3} & \frac{\Delta^{2}}{2} & 0 & 0\\
\frac{\Delta^{2}}{2} & \Delta & 0 & 0\\
0 & 0 & \frac{\Delta^{3}}{3} & \frac{\Delta^{2}}{2}\\
0 & 0 & \frac{\Delta^{2}}{2} & \Delta
\end{smallmatrix}\right]
\end{align*}
is the covariance matrix of the process noise $\boldsymbol{w}_{k}$,
$\Delta$ is the time step size, and the state vector $\boldsymbol{x}_{k}$
denotes the position and velocity components along the coordinate
axes, i.e., $\boldsymbol{x}_{k}=\left[x_{k},\thinspace\dot{x}_{k},\thinspace y_{k},\thinspace\dot{y}_{k}\right]^{T}$.
As shown in Fig.~\ref{fig:Comm-topology}, $50$ sensing agents are distributed over the given region and are able to communicate
with their neighboring agents. The undirected communication network
topology is assumed to be time-invariant. Local-degree weights are
used to compute the doubly stochastic adjacency matrix $\mathcal{A}_{k}$
as:\vspace{-5pt}
\begin{align*}
\mathcal{A}_{k}[i,j] & =\frac{1}{\max(d_{i},d_{j})}\thinspace,\qquad\forall j\in\mathcal{J}_{k}^{i}\textrm{ and }i\not=j\thinspace,\\
\mathcal{A}_{k}[i,i] & =1-\sum_{j\in\mathcal{V}\backslash\{i\}}\mathcal{A}_{k}[i,j]\thinspace,
\end{align*}
where $d_{i}$ denotes the degree of the $i^{\textrm{th}}$ agent.

\begin{figure}[!h]
\begin{centering}
\begin{tabular}{cc}
\includegraphics[width=1.4in]{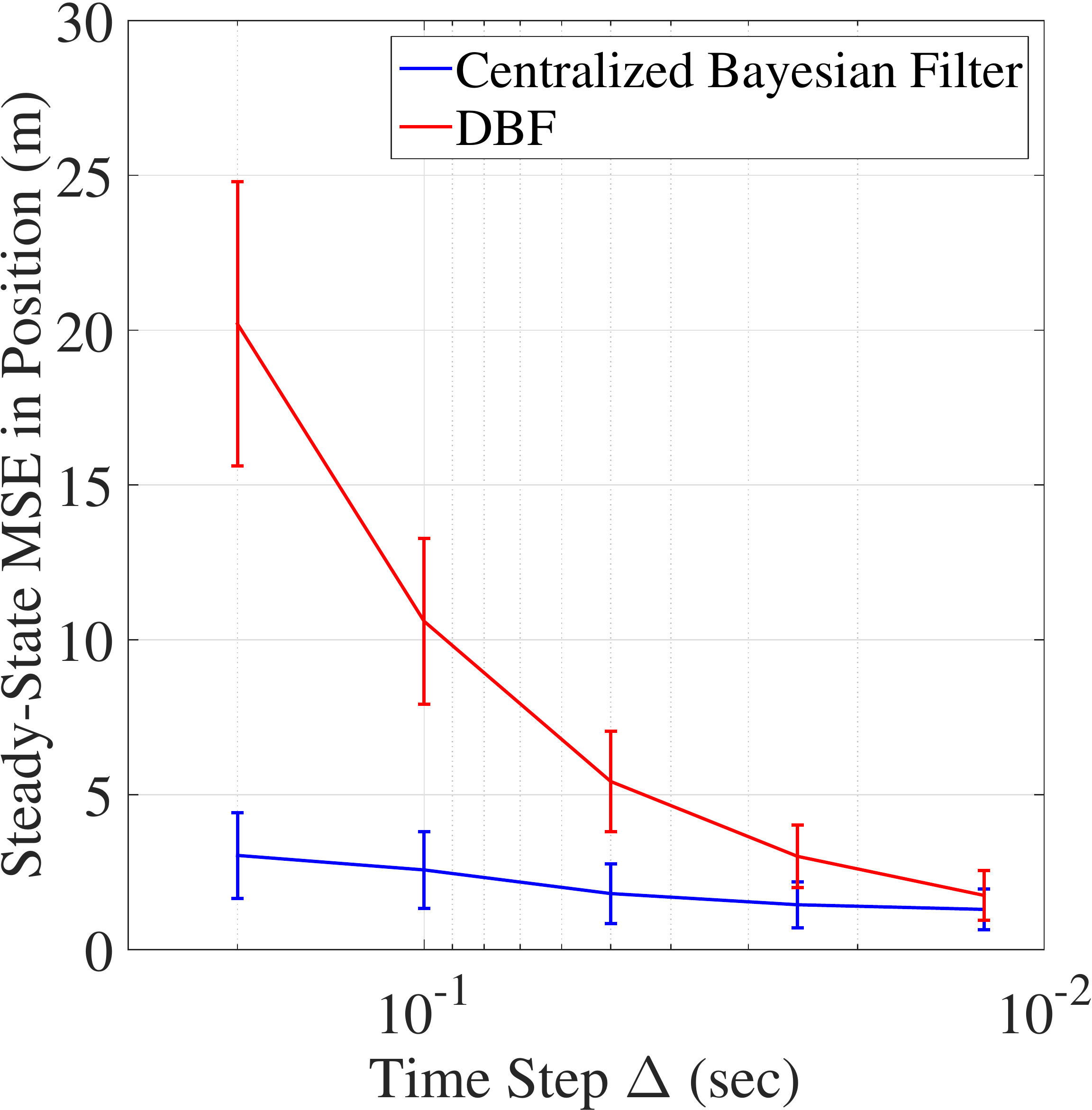} & \includegraphics[width=1.4in]{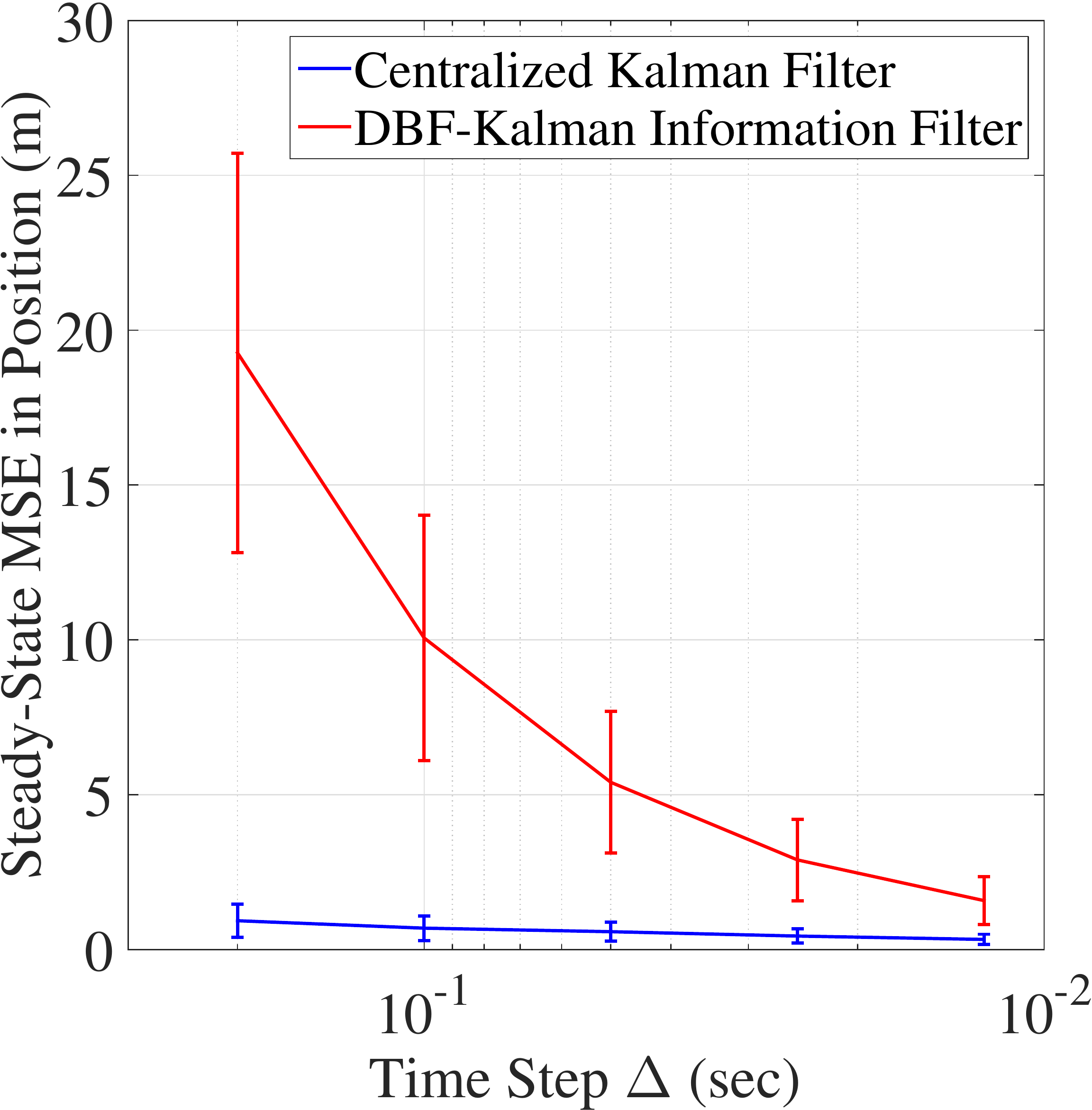}\tabularnewline
(a) & (b)\tabularnewline
\end{tabular}
\par\end{centering}
\caption{Variation of steady-state MSE in position with respect to time step size
$\Delta$ is shown for (a) the centralized Bayesian filtering algorithm
and the DBF algorithm in Scenario 1 and (b) the centralized Kalman
filtering algorithm and the DBF algorithm for linear-Gaussian models
in Scenario 2. \label{fig:SSMSE-errorbar}}
\end{figure}

\begin{figure}[!h]
\begin{centering}
\includegraphics[width=2.5in]{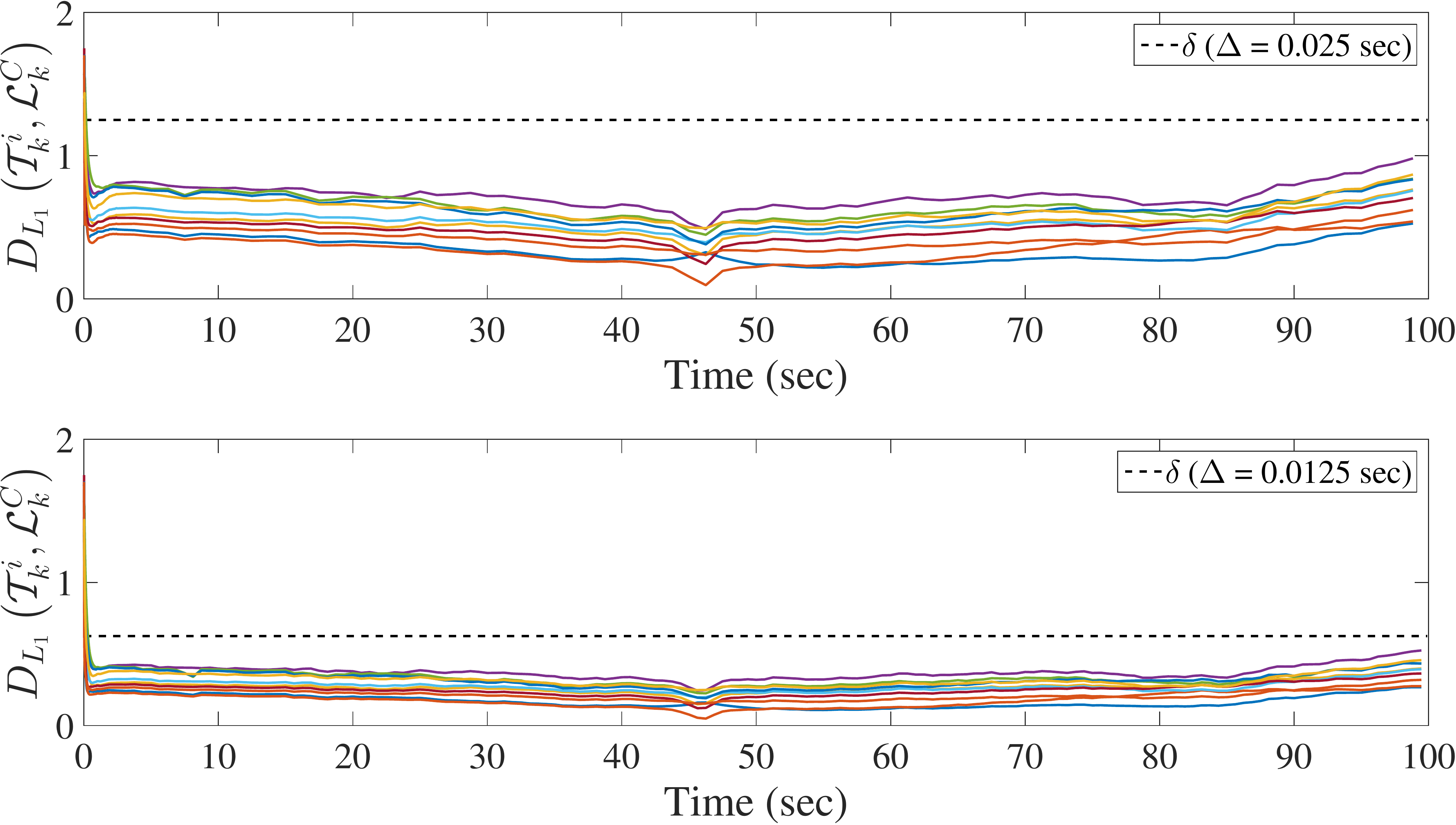}
\par\end{centering}

\caption{The trajectories of the $L_{1}$ distances between the estimated likelihood functions and the joint likelihood function for the ten sensing agents are shown.
\label{fig:DBF-L1-distance}}
\end{figure}

%\subsection{Scenario 1: Nonlinear Measurement Models \label{sub:Scenario-1:-Nonlinear}}
In Scenario~1, five of these agents are equipped with nonlinear
position sensors that can measure their distance to the target using
Time of Arrival (TOA) sensors. Another five agents are equipped with
Direction of Arrival (DOA) sensors that can measure the bearing angle between the
target and themselves. The remaining agents do not have any sensors.
The measurement models for these sensors are given by: \vspace{-5pt}
\begin{align}
 & \boldsymbol{h}_{k}^{i}(\boldsymbol{x}_{k},\boldsymbol{v}_{k}^{i})= \label{eq:benchmark_sensor}\\
 & \begin{cases}
\textrm{atan2}(x_{k}-x^{i},\thinspace y_{k}-y^{i})+\boldsymbol{v}_{k,\mathrm{DOA}}^{i} & \textrm{for DOA sensor}\\
\sqrt{(x_{k}-x^{i})^{2}+(y_{k}-y^{i})^{2}}+\boldsymbol{v}_{k,\mathrm{TOA}}^{i} & \textrm{for TOA sensor}
%\boldsymbol{v}_{k}^{i} & \textrm{if }i\textrm{ has no sensor}
\end{cases}, \nonumber
\end{align}
where $(x^{i},y^{i})$ denotes the position of the $i^{\textrm{th}}$
agent and atan2 is the 4-quadrant inverse tangent function.
The DOA sensor's measurement noise $\boldsymbol{v}_{k,\mathrm{DOA}}^{i} = \mathcal{N}(0,\sigma_{\theta})$ has variance $\sigma_{\theta}=2^{\circ}$ and the TOA sensor's measurement noise $\boldsymbol{v}_{k,\mathrm{TOA}}^{i} = \mathcal{N}(0,\sigma_{r})$ has variance $\sigma_{r}=10$~m.
Each agent executes the DBF algorithm in Algorithm~\ref{alg:DBF} using particle filters with $10^4$ particles. The comparison between the DBF algorithm and the centralized Bayesian filtering algorithm for varying time step sizes ($\Delta$) is shown in Fig.~\ref{fig:SSMSE-errorbar}(a).
The same target motion, shown in Fig.~\ref{fig:Comm-topology}, is used for all simulations.
We see that the DBF algorithm's steady-state mean-square-error (MSE) in position converges to that of the centralized
algorithm as the time step size $\Delta$ decreases (i.e., the steady-state MSE is smaller than $5$ m if the time step size $\Delta \leq 0.05$ sec).
Note that the MSE of the centralized algorithm does not change much with time step size because it is constrained by the measurement noise intensities.
This shows that the performance of the DBF algorithm approaches the performance of the centralized Bayesian filter as the time step size is reduced.
Moreover, Fig.~\ref{fig:DBF-L1-distance} shows that the $L_{1}$ distances between the estimated likelihood functions and the joint likelihood function are bounded by $\delta$.
In Scenario~2, the same ten agents (having DOA or TOA sensors) have linear position sensors $\boldsymbol{h}_{k}^{i}(\boldsymbol{x}_{k},\boldsymbol{v}_{k}^{i})= \left[\begin{smallmatrix}1 & 0 & 0 & 0\\
0 & 0 & 1 & 0
\end{smallmatrix}\right]\boldsymbol{x}_{k}+\boldsymbol{v}_{k,lin}^{i}$,
with measurement noise $\boldsymbol{v}_{k,lin}^{i} = \mathcal{N}( \boldsymbol{0}, R_{k}^{i})$ and covariance matrix $R_{k}^{i}=15\mathbf{I}$.
Here, each agent executes the distributed Kalman information filtering algorithm from Algorithm~\ref{alg:DBF-LGM}.
Fig.~\ref{fig:SSMSE-errorbar}(b) shows that the performance of the distributed Kalman information filtering algorithm approaches the performance of the centralized Kalman filtering algorithm as the time step size is reduced.

\subsection{Relative Position Estimation for Formation\label{sub:Sim-Multiagent-Formation}}
\begin{figure}[!h]
\begin{centering}
\begin{tabular}{cc}
\includegraphics[width=1.2in]{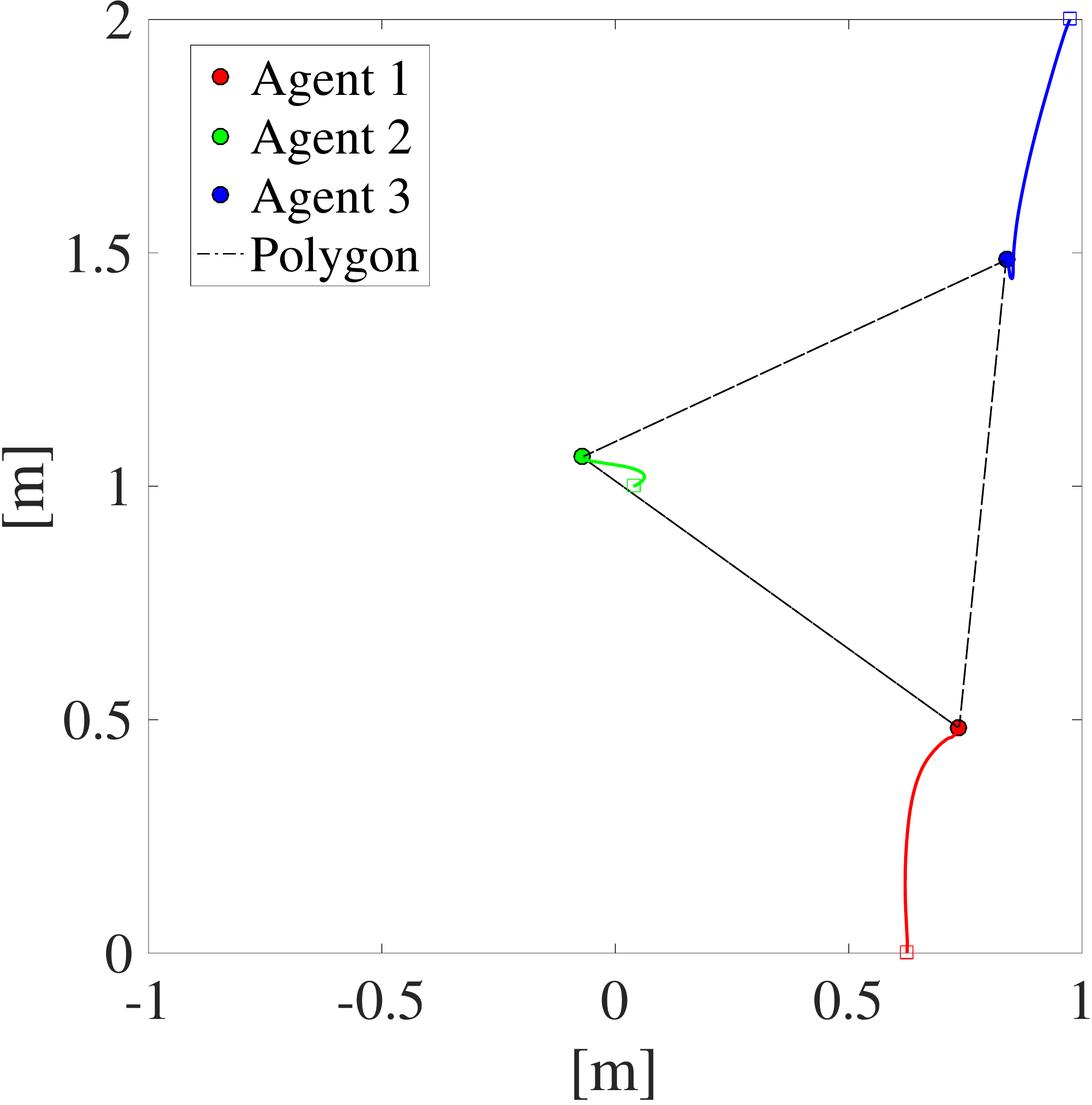} & \includegraphics[width=1.2in]{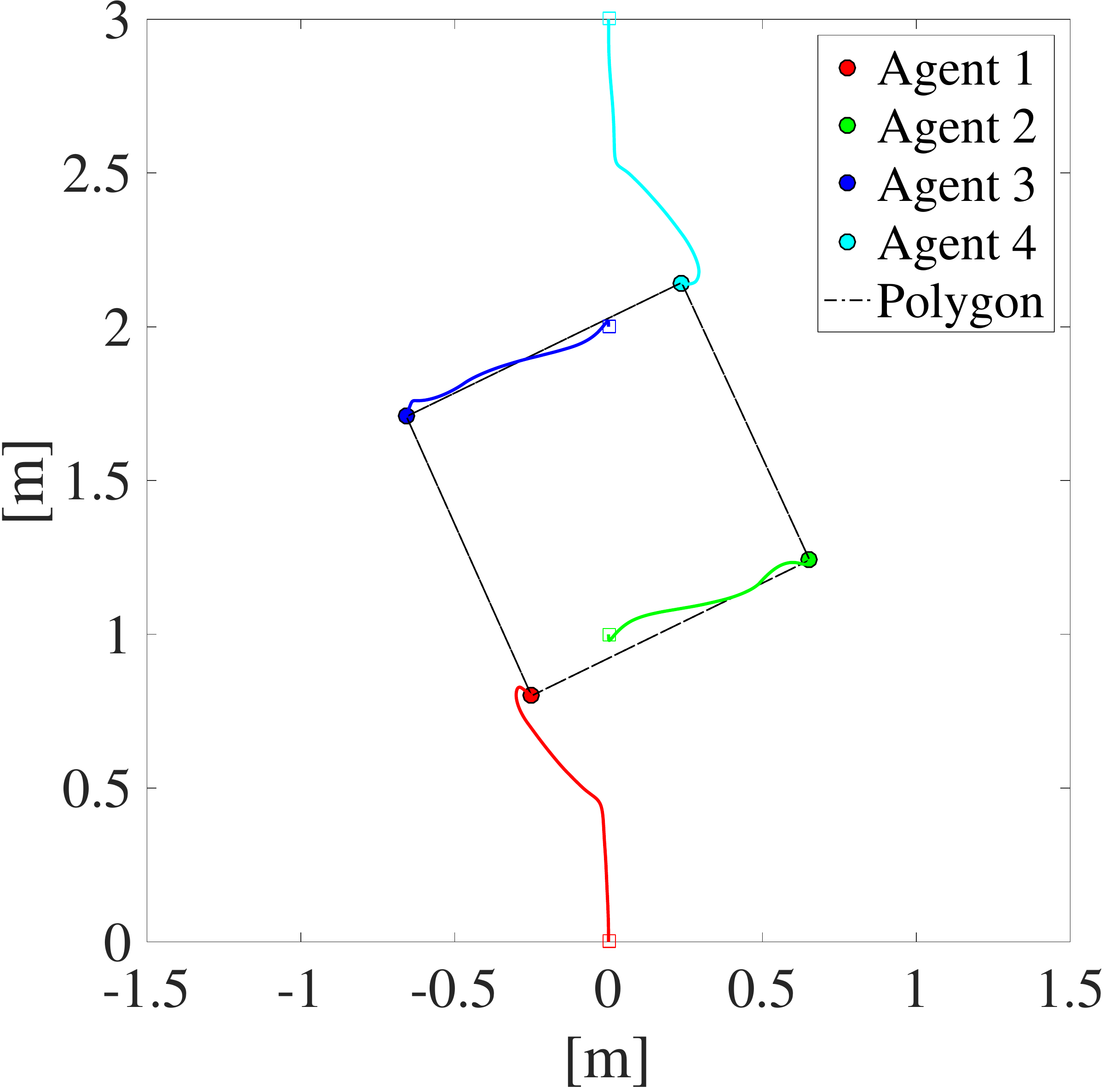}\tabularnewline
(a) $N = 3$ & (b) $N = 4$ \tabularnewline
\includegraphics[width=1.2in]{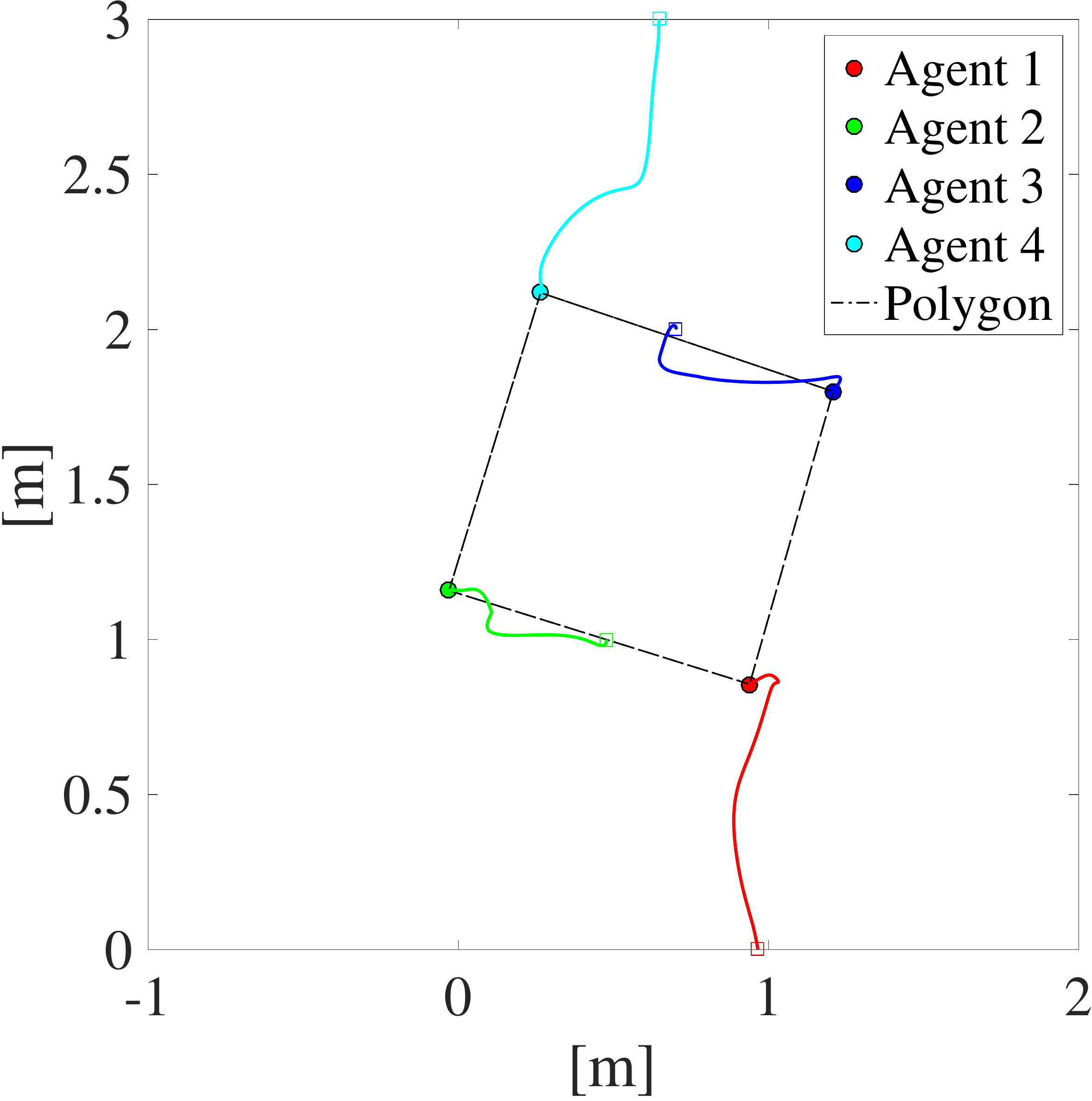} & \includegraphics[width=1.2in]{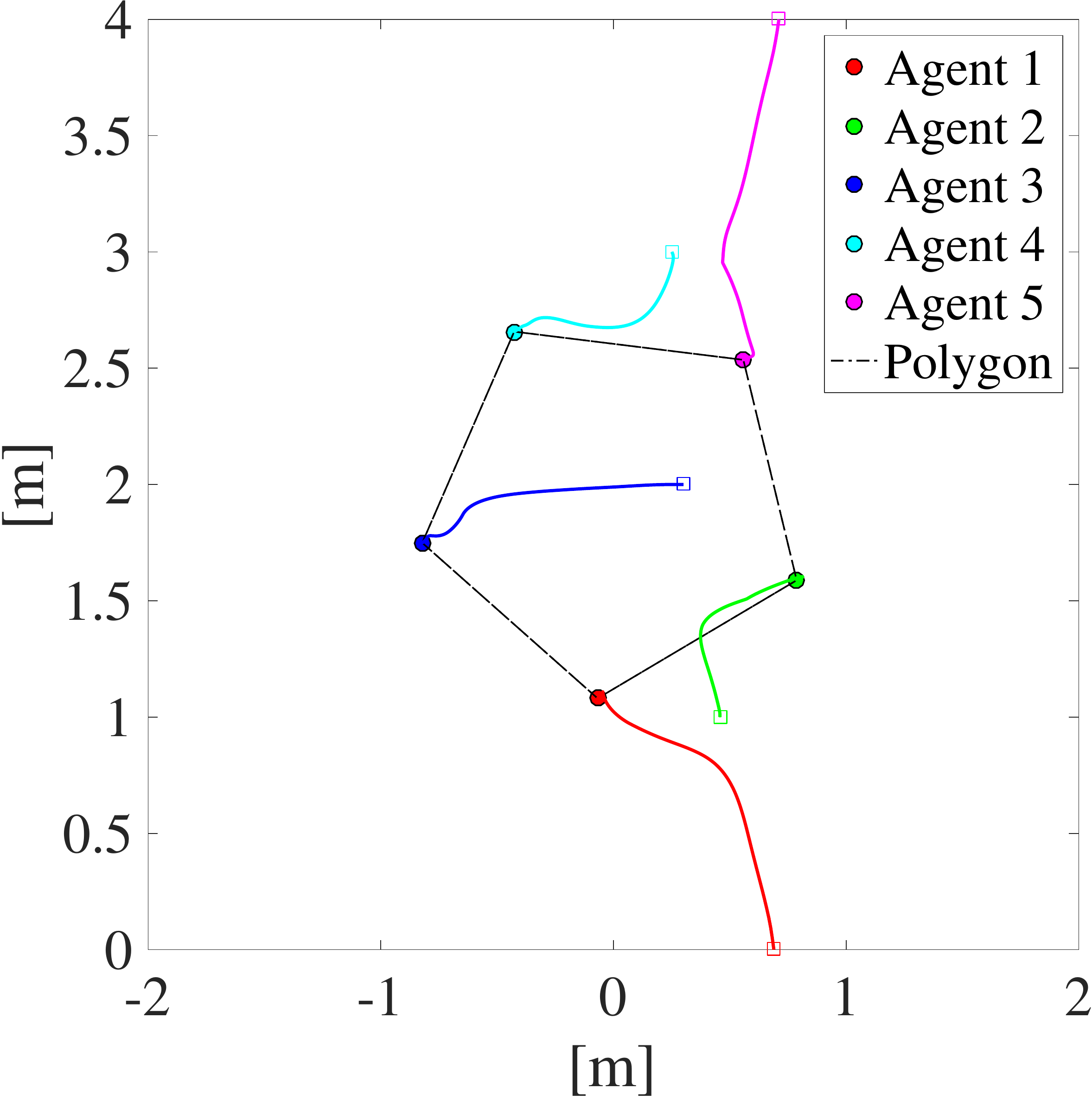}\tabularnewline
(c) $N = 4$ & (d) $N = 5$ \tabularnewline
\end{tabular}
\par\end{centering}
\caption{The initial position ($\square$), the final position (\tikzcircle[fill=none]{3pt}), the trajectories of all the agents, and the final regular polygon are shown for $N=3,4,5$ agents.  \label{fig:maf}}
\end{figure}

In this subsection, $N$ agents estimate their relative positions using only range measurements, and then reconfigure to a $N$-sided regular polygon.
Specifically, each agent can only measure the distance to its nearest two neighbors using a TOA sensor, whose measurement model is described in (\ref{eq:benchmark_sensor}).
Each agent simultaneously executes $N$ DBF algorithms to estimate the relative positions of all the agents.
The $i^\textrm{th}$ agent's dynamics and control inputs are given by: \vspace{-5pt}
\begin{align*}
& \boldsymbol{x}_{k+1}^i = \boldsymbol{x}_{k}^i + \Delta \boldsymbol{u}_{k}^i \thinspace, \\
& \boldsymbol{u}_{k}^i = \sum_{j \in \mathcal{N}_k^i} APF( \hat{\boldsymbol{x}}_{k}^{i,j}, \hat{\boldsymbol{x}}_{k}^{i,i}, d) + APF( \hat{\boldsymbol{x}}_{k}^{i,CM}, \hat{\boldsymbol{x}}_{k}^{i,i}, d_{CM})
\end{align*}
where $\mathcal{N}_k^i$ denotes the two nearest neighbors of the $i^\textrm{th}$ agent and $\hat{\boldsymbol{x}}_{k}^{i,j}$ is the $i^\textrm{th}$ agent's estimate of the $j^\textrm{th}$ agent's position, which is obtained using the DBF algorithms.
The agents use the artificial potential field (APF) based approach to maintain a distance $d$ from their nearest neighbors:
$
APF( \hat{\boldsymbol{x}}_{k}^{i,j}, \hat{\boldsymbol{x}}_{k}^{i,i}, d) = \frac{(\hat{\boldsymbol{x}}_{k}^{i,j} - \hat{\boldsymbol{x}}_{k}^{i,i})} {r_k^{i,j}} \left( a \thinspace r_k^{i,j} - \frac{a \thinspace d^2}{r_k^{i,j}}  \right)$, where $r_k^{i,j} = \| \hat{\boldsymbol{x}}_{k}^{i,j} - \hat{\boldsymbol{x}}_{k}^{i,i} \|_2 $, and maintain a distance $d_{CM} = \frac{d}{2 \thinspace \cos \left( \frac{\pi}{2} - \frac{\pi}{N} \right) }$ from the estimated center of mass $\hat{\boldsymbol{x}}_{k}^{i,CM} = \frac{1}{N} \sum_{j=1}^N \hat{\boldsymbol{x}}_{k}^{i,j}$.
In the propagation step of the DBF algorithm, the agents use their estimated positions to estimate the control input applied by other agents.
Therefore, the estimation errors contribute to the process noise in the propagation step.
During the fusion step at $k^{\textrm{th}}$ time instant, the $i^\textrm{th}$ agent communicates with the $j^\textrm{th}$ agent if either $j \in \mathcal{N}_k^i$ or $i \in \mathcal{N}_k^j$.
In these simulations, we use $a=0.1$, $d = 1$ m, $\Delta = 0.1$ sec, and $10^3$ particles to execute each DBF algorithm.
At the start of the estimation process, the particles are selected from a uniform distribution over the state space $\mathcal{X} = [-N, N] \times [-N, N]$.
The simulation results for multiple values of $N$ are shown in Fig.~\ref{fig:maf}. Since the agents only use relative measurements, the orientation of the final $N$-sided regular polygon in the global frame is not fixed. Therefore, we conclude that the $N$ agents successfully estimate their relative positions using the DBF algorithms and achieve the complex desired formations.
\section{Conclusions \label{sec:Conclusions}}\vspace{-5pt}
In this paper, we presented a novel, discrete-time distributed estimation
algorithm, namely the DBF algorithm, that ensures that each agent's
estimated likelihood function converges to an error ball around the
joint likelihood function of the centralized multi-sensor Bayesian
filtering algorithm. We have rigorously proven the convergence properties
of this algorithm. We have shown an explicit connection between
the time step size of the distributed estimation algorithm and the time-scale
of the target dynamics.
We also presented the distributed Kalman information filtering algorithm for the special case of linear-Gaussian models.
The properties of these algorithms are illustrated
using complex numerical examples. We envisage that the novel
proof techniques presented in this paper can also be used in other
distributed estimation algorithms which rely on the LogOP scheme.

%\begin{ack}
%This research was supported by AFOSR grant \\ FA95501210193 and NSF grant IIS-1253758.
%\end{ack}

%\bibliographystyle{plain} %ieeetr} %plain}
\bibliography{SapBib}
\appendix
\section{Proof of Lemma~\ref{lem:psi-exisits} \label{appendix:proof-lem:psi-exisits}}\vspace{-5pt}
If this claim is untrue, then either $0<\mathcal{P}(\boldsymbol{x})<\mathcal{Q}(\boldsymbol{x})$
or $0<\mathcal{Q}(\boldsymbol{x})<\mathcal{P}(\boldsymbol{x})$ for all $\boldsymbol{x}\in\mathcal{X}$.
% due to Assumption \ref{assump:nonnegative_pdf}.
Hence either $\int_{\mathcal{X}} \mathcal{P}(\boldsymbol{x}) d\mu(\boldsymbol{x})=1 <\int_{\mathcal{X}}\mathcal{Q}(\boldsymbol{x})d\mu(\boldsymbol{x})$
or $\int_{\mathcal{X}}\mathcal{Q}(\boldsymbol{x})d\mu(\boldsymbol{x}) <  \int_{\mathcal{X}} \mathcal{P}(\boldsymbol{x})d\mu(\boldsymbol{x}) = 1$,
which results in contradiction since $\int_{\mathcal{X}}\mathcal{Q}(\boldsymbol{x})d\mu(\boldsymbol{x})=1$.
Hence, such a $\boldsymbol{\psi}\in\mathcal{X}$ must exist.
\section{Proof of Lemma~\ref{lem:conv-H-func-implies-conv-pdf} \label{appendix:proof-lem:conv-H-func-implies-conv-pdf}}\vspace{-5pt}
 Since $\lim_{k\rightarrow\infty}\mathscr{P}_{k}^{i}(\boldsymbol{x})=\mathscr{P}^{\star}(\boldsymbol{x})$, we have \\ $\lim_{k\rightarrow\infty} \! \left(\log\mathcal{P}_{k}^{i}(\boldsymbol{x})\!-\!\log\mathcal{P}_{k}^{i}(\boldsymbol{\psi})\right)  \!=\! \log\mathcal{P}^{\star}(\boldsymbol{x})\!-\!\log\mathcal{P}^{\star}(\boldsymbol{\psi}).$
From Lemma~\ref{lem:psi-exisits}, substituting $\lim_{k\rightarrow\infty}\mathcal{P}_{k}^{i}(\boldsymbol{\psi})=\mathcal{P}^{\star}(\boldsymbol{\psi})$ gives
%$\lim_{k\rightarrow\infty}\ln\mathcal{P}_{k}^{i}(\boldsymbol{x})=\ln\mathcal{P}^{\star}(\boldsymbol{x})$
%for all $\boldsymbol{x}\in\mathcal{X}$. This implies that
$\lim_{k\rightarrow\infty}\mathcal{P}_{k}^{j}(\boldsymbol{x})=\mathcal{P}^{\star}(\boldsymbol{x})$
%for all $\boldsymbol{x}\in\mathcal{X}$
since logarithm is a monotonic
function.
\section{Proof of Lemma~\ref{lem:convegence-TV} \label{appendix:proof-lem:convegence-TV}}\vspace{-5pt}
 It follows from Scheff$\acute{\textrm{e}}$'s theorem \cite[pp. 84]{Ref:Durrett05} that if the pdfs converge pointwise, then their induced measures converge in TV.
The relationship between TV error and $L_{1}$ distance follows from \cite[pp. 48]{Ref:Peres09}.

\end{document}